\documentclass[twocolumn]{aastex63}
\usepackage[utf8]{inputenc}

\usepackage{graphicx}
\usepackage{xcolor}
\usepackage{amsmath}
\usepackage{amssymb}

\usepackage{array}
\newcolumntype{H}{>{\setbox0=\hbox\bgroup}c<{\egroup}@{}}		

\newcommand{\Lir}{\ensuremath{\textrm{L}_{\textrm{IR}}/\textrm{L}_\star}}
\newcommand{\Mjup}{\,\ensuremath{\rm{M_{J}}}}
\newcommand{\kmpers}{\,\mbox{km\,s$^{-1}$}}
\newcommand{\Msun}{\,\ensuremath{\rm{M}_\odot}}

\newcommand{\gaia}{\textit{Gaia}}
\newcommand{\wise}{WISE}
\newcommand{\sphere}{VLT/SPHERE}

\usepackage{hyperref}

\newcommand{\mitmki}{Department of Physics, and Kavli Institute for Astrophysics and Space Research, Massachusetts Institute of Technology, Cambridge, MA02139, USA}
\newcommand{\exe}{University of Exeter, School of Physics \& Astronomy, Stocker Road, Exeter, EX4 4QL, UK}
\newcommand{\lam}{Aix Marseille Univ, CNRS, CNES, LAM, Marseille, France}
\newcommand{\jpl}{Jet Propulsion Laboratory, California Institute of Technology, 4800 Oak Grove drive, Pasadena CA 91109, USA}
\newcommand{\caltech}{Cahill Center for Astronomy and Astrophysics, California Institute of Technology, 1200 East California Boulevard, MC 249-17,Pasadena, CA 91125, USA}
\newcommand{\ipac}{IPAC, California Institute of Technology, M/C 100-22, 1200 East California Boulevard, Pasadena, CA 91125, USA}
\newcommand{\ukansas}{The University of Kansas, Department of Physics and Astronomy, Malott Room 1082, 1251 Wescoe Hall Drive, Lawrence, KS, 66045,USA}
\newcommand{\flatiron}{Center for Computational Astrophysics, Flatiron Institute, New York, NY 10010, USA}
\newcommand{\unige}{Observatoire de l’Universit\'e de Gen\`eve, Chemin Pegasi 51, 1290 Versoix, Switzerland}

\received{...}
\revised{...}
\accepted{...}

\shorttitle{\sphere~survey of dusty \wise~disks}
\shortauthors{Matthews et al.}

\begin{document}

\title{Three new late-type stellar companions to very dusty WISE debris disks identified with VLT/SPHERE imaging}

\correspondingauthor{Elisabeth C. Matthews}
\email{elisabeth.matthews@unige.ch}

\author[0000-0003-0593-1560]{Elisabeth C. Matthews}
\affiliation{\unige}
\affiliation{\mitmki}

\author{Sasha Hinkley}
\affiliation{\exe}

\author{Karl Stapelfeldt}
\affiliation{\jpl}

\author[0000-0002-5902-7828]{Arthur Vigan}
\affiliation{\lam}

\author{Dimitri Mawet}
\affiliation{\caltech}

\author{Ian J. M. Crossfield}
\affiliation{\ukansas}

\author[0000-0001-6534-6246]{Trevor J.\ David}
\affiliation{\flatiron}

\author{Eric Mamajek}
\affiliation{\jpl}

\author[0000-0001-6126-2467]{Tiffany Meshkat}
\affiliation{\ipac}

\author[0000-0001-9414-3851]{Farisa Morales}
\affiliation{\jpl}

\author{Deborah Padgett}
\affiliation{\jpl}


\begin{abstract}

Debris disk stars are good targets for high contrast imaging searches for planetary systems, since debris disks have been shown to have a tentative correlation with giant planets. We selected 20 stars identified as debris disk hosts by the WSIE mission, with particularly high levels of warm dust. We observed these with the VLT/SPHERE high contrast imaging instrument with the goal of finding planets and imaging the disks in scattered light. Our survey reaches a median 5$\sigma$~sensitivity of 10.4\Mjup~at 25au and 5.9\Mjup~at 100au. We identified three new stellar companions (HD18378B, HD19257B and HD133778B): two are mid-M type stars and one is late-K or early-M star. Three additional stars have very widely separated stellar companions (all at $>$2000au) identified in the \textit{Gaia} catalog. The stars hosting the three SPHERE-identified companions are all older ($\gtrsim$700Myr), with one having recently left the main sequence and one a giant star. We infer that the high volumes of dust observed around these stars might have been caused by a recent collision between the planets and planetesimal belts in the system, although for the most evolved star, mass loss could also be responsible for the infrared excess. Future mid-IR spectroscopy or polarimetric imaging may allow the positions and spatial extent of these dust belts to be constrained, thereby providing evidence as to the true cause of the elevated levels of dust around these old systems. None of the disks in this survey are resolved in scattered light.
\vspace{1cm}

\end{abstract}


\section{Introduction}

Many nearby stars older than a few Myr host debris disks, the second-generation belts of dust produced by collisional grinding of planetesimals in the system \citep[for reviews see][]{wyatt2008,matthews2014}. It has been theorized that these disks are correlated with the presence of giant, wide-separation planets. Small dust grains are rapidly removed by stellar winds, and so the very existence of a debris belt implies that there are also larger planetesimals in the system, that are experiencing ongoing collisions to replenish the dust. Not only are these planetesimals the building blocks of planets, but the presence of one or more substellar companions in such a system would perturb the orbits of these planetesimals, enhancing the rate of collisions and thus increasing the rate of dust production. To this end, a link between debris disks and giant planets on long orbits has long been theorized \citep[e.g.][]{mustillwyatt2009} and observationally sought after \citep[e.g.][]{wahhaj2013,rameau2013,janson2013,meshkat2017}.

Debris disks are typically inferred based on excess infrared emission above the stellar photosphere of the target in question, with the temperature and fractional luminosity of this infrared emission indicating the radius and volume of the dust, respectively, although there are degeneracies between dust radius and composition \citep[e.g.][]{pawellek2015}. Several missions have searched for debris disk stars based on their IR excess: in particular, the all-sky \wise~survey \citep{wright2010} surveyed the sky in four near- and mid-IR filters. This has led to many new targets with high volumes of circumstellar dust being identified \citep[e.g.][]{padgett2011,cottensong2016}, which are well suited to direct imaging searches.

The relationship between debris disks and directly imageable planetary companions is hard to quantify observationally, due to the inherent difficulty in detecting these planetary companions with current instruments.  Nonetheless, \citet{meshkat2017} demonstrated a tentative correlation between the two populations: a statistically different occurrence rate was observed for planets in debris disk systems versus those in a control sample, at the 88\% confidence level, with $6.3^{+3.5}_{-2.6}\%$ of debris disk systems hosting wide separation giant planets, compared to just $0.7^{+1.1}_{-0.5}\%$ of those without debris disks. To strengthen this relation requires additional deep observations of nearby dusty systems, so as to increase the available sample size of highly dusty systems with constraints on their giant planet populations. Further, the theoretical and observational suggestions that planets are more common in dusty systems makes these dusty systems an excellent place to search for giant planets with direct imaging, since the likelihood of detections is elevated by the higher occurrence rate.

Warm debris disks are an indicator of youth. These disks have temperatures of a few tens to a few hundreds of kelvin, and can be detected via a 12$\mu$m or 22$\mu$m infrared excess. \citet{wyatt2007a} predicted that disk luminosity should fall as $t^{-1}$, and there is strong observational evidence that debris disk occurrence rates decrease with age, a result that holds across spectral type \citep[e.g.][]{rieke2005,su2006,carpenter2009,plavchan2009,thureau2014,theissenwest2014}. \citet{kennedywyatt2013} find that only 1 in 10,000 old stars ($>$Gyr) have a 12$\mu$m excess, whereas 1 in 100 young stars ($<$120Myr) possess such an infrared excess. A survey targeting debris disk stars is therefore biased towards young targets, even if no explicit age selection is performed. Young planets are bright, due to the re-radiation of the energy liberated during formation, and so are easier to detect than old planets at the same mass. Debris disk hosts are prime targets for direct imaging: the occurrence rate of giant, wide-separation planets in these systems is likely elevated, and these targets are typically young and therefore host bright planets.

These targets are also potentially amenable to direct detection of the debris disks in scattered light. In these cases, many opportunities to directly characterize the disk are available: spectroscopic and photometric measurements may reveal the composition of the disk, while the fractional polarization and scattering phase function allow the grain structure and porosity to be constrained \citep[e.g][]{stark2014,schneider2014,perrin2015,milli2017}. Dynamical modeling can also allow additional constraints on the likely positions and masses of planets in the system \citep[e.g.][]{mouillet1997,wyatt2005a,chiang2009,nesvoldkuchner2015,leechiang2016,shannon2016}.

Dust is also occasionally detected around older stars, sometimes in remarkably high quantities. \citet{theissenwest2017} identified extreme mid-infrared excesses (\Lir$>$0.01) around a number of M-dwarfs, with ages $>$1Gyr. Typical dust dispersal timescales are tens to hundreds of Myr, suggesting that this excess is not the remnants of primordial dust but instead likely the result of dramatic collisions later in the lifetime of the planetary system. \citet{weinberger2011} studied in detail the late-F spectroscopic binary BD+20 307, an old ($\gtrsim$1Gyr) star that hosts an unusually large quantity of dust. The system lacks cold dust, implying that the dust cannot be caused in a collisional cascade \citep{wyatt2007b,hengtremaine2010} or in a process akin to the late heavy bombardment \citep{gomes2005,strom2005}. Instead, \citet{weinberger2011} infer that the dust was likely produced in a cataclysmic collision when a terrestrial planet was destabilized from its orbit. They calculate that such collisions are relatively common, with AFG-type stars hosting planetary systems that undergo at least 0.2 impacts per main-sequence star lifetime. Similarly the HD69830 system has significant dust quantities at an age of $>$1Gyr \citep{beichman2005}, and is orbited by at least three Neptune-sized planets \citep{lovis2006}. Perhaps those planets contributed to the destabilization of a terrestrial planet, causing a collision and thereby an elevated quantity of dust in the system.  HD23514 \citep{rhee2008} and HD169666 \citep{moor2009} also host more dust than is reasonable from a collisional cascade, implying that the dust is instead produced in a terrestrial planet collision. On the other hand, \citet{bonsor2013} detect a disk around the sub-giant star $\kappa$ CrB, a system also known to host at least one planet \citep{johnson2008}, and infer the disk likely survived the entire $\sim$2.5Gyr lifetime of the system. These old systems with debris disks allow a unique insight into the evolution and the longevity of planetary systems. In particular, the subset of systems in which the observed dust is the aftermath of a cataclysmic collision opens up opportunities to study the dust properties, and thus infer the bulk compositions of the parent bodies responsible for producing this dust.

In this paper, we present a survey of 20 highly dusty systems identified as having a 22$\mu$m excess, indicating a corresponding belt of warm debris, with the \wise~mission \citep{wright2010}. We chose targets with a high mid-IR excess, that are early type and nearby. We generally avoided targets that had previously been observed with high contrast imaging platforms, and avoided observing known binaries. We did not carry out an explicit age selection, but relied on the correlation between dust and youth to bias the sample towards young stars. Each target was observed with the \sphere~high contrast imager, with the aim of detecting planetary companions and debris disks in scattered light. We detected late-type stellar companions to three of the targets in this survey, and determined that several of the targets we studied were in fact old stars. Indeed, all three stellar companions detected in this work orbit old primary stars, with one having recently left the main sequence and one being a giant star. We did not detect any substellar companions, or resolve any of the debris disks in scattered light.

We describe our target selection process and the dusty nature of these targets in Section \ref{sec:targets}. The observations and data processing are described in Sections \ref{sec:obs} and \ref{sec:dr} respectively. Section \ref{sec:methods} describes our age determination, and astrometric, spectroscopic and photometric analysis of candidate companions, and in Section \ref{sec:companions} we present the three new stellar companions identified with \sphere. Finally, Section \ref{sec:discussion} discusses these results in context.


\section{Target Selection}
\label{sec:targets}

For this work the aim was to select targets with high volumes of circumstellar dust from the all-sky \wise~catalog of sources \citep{cutri2012}. We observed targets that \citet{padgett2011} had identified as debris disk stars. That work selected targets that have:

\begin{enumerate}
	\item A large W1-W4 excess (i.e. [3.6$\mu$m]-[24$\mu$m] color). Many of the targets here have been subsequently observed with Herschel and/or Spitzer, allowing for more detailed characterization of the infrared excess. These additional measurements are not taken into account in our target selection. However, we use fitted values of the disk $\Lir$~and T$_\textrm{eff}$ from \citet{cottensong2016}, who do take into account these additional infrared measurements when calculating the disk properties.
	\item A small offset of less that 0.2'' between the {\it Hipparcos} and \wise~source positions when taking into account the proper motion of the target, so as to ensure that the excess is indeed associated with the foreground star.
	\item Either a Hipparcos-measured distance within 120pc or, if the object was not in Hipparcos, a large proper motion of $>$30mas/yr as a proxy for proximity of the target. The distance cutoff was chosen so as to primarily select debris disks rather than protoplanetary disks. There are several star-forming regions at distances of $\sim$150pc, which include many protoplanetary disks with large W1-W4 excesses. A secondary reason for the distance cut-off is to ensure that targets will be bright and close enough that we are able to reach high contrasts at small physical separations. The 30mas/yr proper motion measurement is typical of stars at a distance of $\sim$120pc. This proper motion cutoff also ensures that background stars and co-moving companions can also easily be differentiated based on their relative motion when two observations are collected a year apart. \citet{padgett2011} also inspect the observations of each candidate visually, and select only those targets that are round and not nebulous. Several of the sources from \citet{padgett2011} were subsequently also identified as debris disk hosts in \citet{cottensong2016}, and a few were mentioned in \citet{wu2013} and \citet{patel2014}, both of which include a more severe distance cut.
\end{enumerate}

We down-selected early-type stars that were visible from Cerro Paranal, where SPHERE is located, from the list of \citet{padgett2011} targets. We also preferentially selected nearby targets. At the time of selection one target (HD18378) had no parallax measurement, but was nonetheless selected based on its large W4 excess and significant proper motion of 37.5mas/yr, suggesting that the target is nearby. This target was subsequently determined by \gaia~to be at a distance of $520.7\pm5.4$pc \citep{gaia_dr2}, which is significantly further than we had anticipated. This target hosts one of the stellar companions we detected, and is discussed in more detail in Section \ref{sec:18378}. 

It is well known that searches for debris disks can be plagued with false positives, such as background galaxies and blended late-type stars, when only color and magnitude cuts are used to select targets \citep[e.g][]{kuchner2016}. However, the visual inspection process carried out by \citet{padgett2011} and repeated in all the surveys mentioned here has been shown to reduce the false positive rate to $\sim$7\% \citep{silverberg2018}. Note also that this contamination rate decreases for brighter excesses due to the brightness distribution of background galaxies, and the targets in this work should therefore have an even lower contamination rate. As such we would expect at most one or two of the targets in this survey to be contaminated, with the remainder being bona fide debris disk stars. 

The final target list is given in Table \ref{tab:dustytargets}. In this table we also include the \wise~W1-W4 excess, representative of the quantity of dust around each object, and list the $\Lir$ value from \citep{cottensong2016} for the 17 targets in that work. \citet{cottensong2016} includes all available infrared photometry to fit the excess, meaning that Herschel and Spitzer observations are used in addition to WISE measurements when available. For this survey we observed 20 objects, with 12 of these subsequently re-observed so as to search for common proper motion of the candidate companions. All of the targets have infrared luminosity $\Lir > 2\times10^{-4}$, with the median distance and stellar type of our sample being 104pc and A1 respectively. No explicit selection was made based on the target ages, but we note that young stars tend to have a larger infrared excess than old targets, as described above, and so we expected that the survey would be biased towards young stars. However, we found that several of the targets have older ages, with a few even having left the main sequence. Section \ref{sec:ages} includes a full discussion of the age determination, and Section \ref{sec:discussion} describes our interpretation of these older debris disk stars.

\begin{table*}[ht]
    \centering
    \footnotesize
    \begin{tabular}{ccHHcccccccc}
        \hline
        HD & HIP & RA(ICRS) & Dec(ICRS) & SpT & SpT Ref & Distance[pc] & W1-W4[mag] & $\Lir$ [10$^{-4}$] & Age[Myr] & Age Quality & Age Ref \\
        \hline \hline
        18378  & ---   & 02:54:13.578 & -62:14:03.85 & K2III    & HC75 & $520.7\pm5.4$ & $1.27\pm0.09$ & ---  & --- & --- & --- \\
        19257  & 14479 & 03:06:51.913 & +30:31:37.41 & F0V      & K16  & $74.4\pm0.3$ & $2.10\pm0.08$  & 3.7  &  785$\pm$146 & 1 & This Work \\
        24966  & 18437 & 03:56:29.376 & -38:57:43.81 & A0V      & H82  & $114.9\pm0.5$ & $1.73\pm0.07$ & 10.9 & 128$^{+99}_{-77}$ & 2 & N13 \\
        94893  & 53484 & 10:56:29.495 & -48:19:55.26 & F0V      & H78  & $106.5\pm0.4$ & $2.21\pm0.04$ & 9.2  & 735$^{+719}_{-313}$ & 2 & DH15 \\
        98363  & 55188 & 11:17:58.137 & -64:02:33.35 & A2V	    & HC75 & $138.6\pm0.7$ & $2.73\pm0.03$ & 1.8  & 17$\pm$8 & 3 & DZ99, H00, R11 \\
        113902 & 64053 & 13:07:38.284 & -53:27:35.15 & B8V      & E66  & $100.4\pm1.2$ & $0.81\pm0.13$ & 10.5 & 16$\pm$8 & 3 & DZ99, H00, R11 \\
        119152 & 66837 & 13:41:52.778 & -18:59:05.44 & F0V      & HS88 & $81.2\pm0.3$ & $0.62\pm0.08$  & 8.2  &  613$^{+867}_{-297}$ & 2 & DH15 \\
        122802 & ---   & 12:59:09.200 & +07:54:02.73 & F3/5V    & HS88 & $105.6\pm0.9$ & $1.49\pm0.05$ & ---  & 280$\pm$224 & 1 & This Work \\
        123247 & 69011 & 14:07:40.807 & -48:42:14.49 & B9.5V    & H78  & $98.8\pm0.7$  & $2.08\pm0.05$ & 21   & 16$\pm$8 & 3 & H00, R11 \\
        133778 & 73976 & 15:07:06.335 & -28:49:23.10 & F3/5V    & HS88 & $163.6\pm1.9$ & $0.89\pm0.08$ & 7.2  & 1480$^{+210}_{-180}$ & 1 & C11 \\
        138564 & 76234 & 15:34:20.853 & -39:20:57.14 & B9V      & HS88 & $102.6\pm0.7$ & $1.15\pm0.08$ & 14.2 & 16$\pm$8 & 3 & H00, R11 \\
        138923 & 76395 & 15:36:11.363 & -33:05:34.10 & B8V      & H70  & $134.2\pm2.2$ & $1.32\pm0.27$ & 6.6  & 16$\pm$8 & 3 & DZ99, H00, R11 \\
        151012 & 82069 & 16:45:48.463 & -26:38:57.90 & B9.5V    & HS88 & $108.6\pm0.6$ & $1.10\pm0.08$ & 10.2 & 16$\pm$8 & 3 & R11 \\
        151029 & 81971 & 16:44:42.410 & +02:19:58.28 & A5V      & AC00 & $109.5\pm0.6$ & $0.86\pm0.08$ & 13.3 & 97$\pm$9 & 1 & This Work \\
        157728 & 85157 & 17:24:06.587 & +22:57:37.02 & A7V      & AM95 & $42.8\pm0.1$  & $1.16\pm0.18$ & 5.1  & 807$^{+301}_{-174}$ & 2 & DH15 \\
        158815 & ---   & 17:32:07.433 & -21:13:10.11 & G1V      & HS88 & $144.3\pm1.2$ & $1.39\pm0.08$ & 2.7  & 1698$\pm$645 & 1 & This Work \\
        176638 & 93542 & 19:03:06.877 & -42:05:42.39 & B9.5Vann & GG87 & $57.5\pm1.2$  & $1.00\pm0.21$ & 37.9 & 93$^{+106}_{-35}$ & 2 & DH15 \\
        182681 & 95619 & 19:26:56.483 & -29:44:35.62 & B8.5V    & H70  & $71.4\pm0.7$  & $1.10\pm0.17$ & 47   & 23$\pm$3 & 3 & G18 \\
        192544 & 99892 & 20:16:02.790 & -16:17:44.47 & A0III    & HS88 & $81.6\pm0.3$  & $0.82\pm0.09$ & ---  & 48$\pm$20 & 1 & This Work \\
        207204 & 107585 & 21:47:24.810 & -04:36:31.13 & A0V     & HS99 & $95.3\pm1.4$  & $0.90\pm0.09$ & 10   & 54$\pm$30 & 1 & This Work \\
        \hline
    \end{tabular}
    \caption{Target stars observed in this survey. Distances are from \gaia~data release 2 \citep{gaia_dr1b,gaia_dr2,lindegren2018}. W1-W4 ([3.6$\mu$m]-[24$\mu$m]) excesses are calculated from the \wise~all-sky survey catalog \citep{cutri2012}, and $\Lir$ values from \citep{cottensong2016} are given where available. We determine ages from the literature where available, and using the \texttt{isochrones} package \citep{mortonisochrones} otherwise (see Section \ref{sec:ages}). References in the table indicate ages for individual targets adopted in our analysis or moving group memberships; for Sco-Cen subgroups we use the ages calculated in \citet{pecautmamajek2016} and for the $\beta$-Pictoris moving group we use the the age from \citet{mamajekbell2014}. To indicate the reliability of each age we additionally assign an age quality flag between 1 and 3, with 3 being the most reliable. The age determinations and quality flag designations are described in detail in Section \ref{sec:ages}. Spectral Type References are AC00: \citealt{abt2000}; AM95: \citealt{abt1995}; E66: \citealt{evans1966}; GG87: \citealt{gray1987}; H70: \citealt{hube1970}; H78: \citealt{houk1978}; H82: \citealt{houk1982}; HC75: \citealt{houk1975}; HS88: \citealt{houk1988}; HS99: \citealt{houk1999}; K16: \citealt{kuchner2016}. Age References are DH15: \citealt{davidhillenbrand2015}; DZ99: \citealt{dezeeuw1999}; G18: \citealt{gagne2018}; H00: \citealt{hoogerwerf2000}; N13: \citealt{nielsen2013}; R11: \citealt{rizzuto2011}. }
    \label{tab:dustytargets}
\end{table*}


\section{Observations}
\label{sec:obs}

We observed each target using the SPHERE instrument at the VLT \citep{beuzit2019}. We used the IRDIFS configuration, which allows simultaneous imaging with both the IRDIS and IFS subsystems (see \citealt{dohlen2008}, \citealt{claudi2008}). The IRDIS subsystem provides imaging over a relatively large field of view (11''$\times$12.5''), while the IFS instrument provides spectra for a smaller field of view (1.73''$\times$1.73'', R$\sim$50). IRDIS was used in dual-band imaging mode \citep{vigan2010} with the H23 filter pair ($\lambda$\,=\,1588.8\,nm, $\Delta \lambda$\,=\,53.1\,nm and $\lambda$\,=\,1667.1\,nm, $\Delta \lambda$\,=\,55.6\,nm), while the IFS was used in the YJ mode, which covers $\lambda$=0.95-1.35$\mu$m and has 39 distinct wavelength channels \citep{zurlo2014,mesa2015}. Our data were collected with the apodized pupil Lyot coronagraph \citep{carbillet2011,guerri2011} in its \texttt{N\_ALC\_YJH\_S} configuration optimized for the H band, allowing for an inner working angle of $\sim$0.15$^{\prime\prime}$.

We observed each target for approximately 2200s in a series of short individual exposures with duration between 8s and 64s. Exact exposure times and number of images for each observation are given in Table \ref{tab:obs}. These images were collected in an angular differential imaging sequence (ADI, \citealt{marois2006}), by using the instrument in pupil-stabilized mode. The multi-wavelength nature of both subsystems means the images also use the spectral differential imaging technique (SDI, \citealt{racine1999}), and the angular and spectral variation of speckles are both used in our analysis, as described in Section \ref{sec:dr}. The average field rotation was 31$^\circ$ for IRDIS data and 33$^\circ$ for IFS data. We collected flux calibration frames, where the position of the target is offset from the coronagraph, and star position calibration frames, where a sinusoidal deformation to the deformable mirror creates four offset copies of the star that can together be used to determine the position of the star behind the coronagraph. These flux and position calibration frames were collected either immediately before or immediately after the science sequence. We additionally used daytime calibrations for flat-field correction.

For 12 of the targets we collected a second epoch of data, so that astrometric monitoring could be used to differentiate between co-moving companions and stationary background objects. These observation sequences were typically shorter, with a median total integration time of 1024s, and had correspondingly smaller field rotations. Exposure times and field rotation for these follow-up observations are also given in Table \ref{tab:obs}.


\section{Data Reduction}
\label{sec:dr}

Data reduction was performed following \citet{matthews2018} and \citet{vigan2015sirius}. We first performed basic calibrations using the ESO data reduction and handling pipeline (DRH, \citealt{pavlov2008}) and a custom IDL pipeline (now available in a python implementation in the \texttt{vlt-sphere}\footnote{see https://github.com/avigan/SPHERE} package, see \citealt{pysphere}). We created master dark and flat frames, and calibrated the star position behind the coronagraph. Frames were then normalized by integration time. In the case of IFS data, we additionally created an IFU flat field, and performed a cross-talk correction and a wavelength calibration correction as described in \citet{vigan2015sirius}. Frames were then realigned based on the star center position calibration and dither position for each frame. Finally, the data were converted from spectra into (x,y,$\lambda$) cubes using the DRH.

We then carried out a principal component analysis \citep[PCA, see e.g.][]{soummer2012,amaraquanz2012}) to remove stellar speckle noise, for both the IRDIS and IFS datacubes. We used a full-frame PCA treatment, where we simultaneously exploit the angular and spectral variation of the speckle noise across the sequence of images \citep{marois2006,racine1999}. We first corrected for the anamorphic distortion of the instrument optics and applied the epsilon correction detailed in the ESO SPHERE User Manual\footnote{https://www.eso.org/sci/facilities/paranal/instruments/sphere/doc.html} which accounts for a slight mis-synchronization between the telescope rotation angle and the header values for each file. We then rescaled the images in proportion to their wavelength, such that the characteristic scale of speckles (proportional to $\lambda / D$, \citealt{racine1999}) was equivalent between images, and on-sky candidate companion signals were physically displaced between images at different wavelengths. The nature of the ADI observing sequence means that stellar speckles are also aligned between timesteps in the original image orientation, while on-sky candidates move their position between frames due to the apparent rotation of the sky. We then used a custom PCA code to find principal components that can be subtracted off: this method selects for similar features between images, and so allows the majority of speckle noise to be captured and removed. For this process we considered the rotational and spectral variation simultaneously. Finally, the images were co-added across time and optionally also across wavelength. Our final data products are both a single broadband image with optimal sensitivity to companions, and a set of narrow-band images where the relative H2 and H3 magnitudes can be measured in the case of IRDIS data, and spectral extraction can be performed in the case of IFS data.

In this process we aimed to optimize the balance of removing sufficient starlight without subtracting too much planetary flux. To do this, we calculated and stored reductions with a range of between 0 and $\sim$1/3 of the available principal components removed. We searched for candidates in several reductions with different numbers of components removed, and when calculating contrast limits took the entire range of reductions into account, as described in Section \ref{sec:ccs}.

\begin{table*}[ht]
    \centering
    \footnotesize
    \begin{tabular}{ll|llHl|llHl}
         & & \multicolumn{4}{c|}{IRDIS} & \multicolumn{4}{c}{IFS} \\
        Target & UT Date & $N_{\textrm{im}}$ & T$_\textrm{exp}$[s] & (red) & Rot[$^{\circ}$] & $N_{\textrm{im}}$ & T$_\textrm{exp}$[s] & (red) & Rot[$^{\circ}$] \\
        \hline
        \hline
        HD 18378  & 2016 Sep 15 &  80 & 32 & 15.7367 & 15.7 &  40 & 64 & 16.5259 & 16.5 \\
                  & 2017 Aug 29 &  32 & 64 & 12.5502 & 12.6 &  32 & 64 & 13.3940 & 13.4 \\
        HD 19257  & 2015 Aug 26 &  48 & 32 &  8.0446 &  8.0 &  49 & 32 &  8.3759 &  8.4 \\
                  & 2017 Sep 01 &  32 & 16 &  2.6631 &  2.7 &  34 & 16 &  3.0055 &  3.0 \\
        HD 24966 & 2016 Sep 19 &  70 & 32 & 32.6106 & 32.6 &  70 & 32 & 33.9566 & 34.0 \\
        HD 94893  & 2015 Apr 11 &  48 & 32 & 16.8792 & 16.9 &  49 & 32 & 17.5337 & 17.5 \\
                  & 2018 Jan 07 &  16 & 16 & 3.07846 &  3.1 &  17 & 16 & 3.44273 &  3.4 \\
        HD 98363  & 2015 Apr 11 &  48 & 32 & 11.4683 & 11.5 &  45 & 32 & 10.9527 & 11.0 \\
                  & 2018 Mar 24 &  16 & 64 &  6.2737 &  6.3 &  16 & 64 &  6.6634 &  6.7 \\
        HD 113902 & 2016 Jun 06 &  80 & 32 & 19.4097 & 19.4 &  80 & 32 & 20.951  & 21.0 \\
                  & 2017 Jun 23 &  64 & 32 & 15.8975 & 15.9 &  32 & 64 & 16.9081 & 16.9 \\
        HD 119152 & 2016 May 01 & 144 & 16 & 48.2286 & 48.2 &  72 & 32 & 53.8343 & 53.8 \\
        HD 122802 & 2016 Jul 25 & 288 &  8 &  9.0661 &  9.1 & 144 & 16 & 10.5613 & 10.6 \\
                  & 2018 May 15 & 192 &  8 & 26.8151 & 26.8 & 180 &  8 & 27.3489 & 27.3 \\
        HD 123247 & 2015 Apr 06 &  48 & 32 & 17.4695 & 17.5 &  49 & 32 & 18.1360 & 18.1 \\
                  & 2018 Mar 22 &  32 & 32 &  9.5631 &  9.6 &  16 & 64 & 10.1635 & 10.2 \\
        HD 133778 & 2016 May 16 & 320 &  8 & 91.3030 & 91.3 & 160 & 16 & 97.8132 & 97.8 \\
                  & 2017 Jun 22 & 256 &  2 & 28.4686 & 28.5 & 136 &  4 & 30.7281 & 30.7 \\
        HD 138564 & 2016 May 07 & 144 & 16 & 33.9917 & 34.0 &  72 & 32 & 35.6218 & 35.6 \\
                  & 2018 Mar 17 &  64 & 16 & 13.4652 & 13.5 &  32 & 32 & 14.3876 & 14.4 \\
        HD 138923 & 2016 May 03 & 144 & 16 & 52.9922 & 53.0 &  72 & 32 & 56.0756 & 56.1 \\
        HD 151012 & 2016 Jun 04 & 320 &  8 & 21.0358 & 21.0 & 160 & 16 & 25.4531 & 25.5 \\
                  & 2017 Jul 15 & 512 &  2 & 13.9968 & 14.0 & 396 &  2 & 15.5371 & 15.5 \\
        HD 151029 & 2016 May 17 &  80 & 32 & 20.8812 & 20.9 &  80 & 32 & 22.5313 & 22.5 \\
        HD 157728 & 2016 May 01 & 160 & 16 & 13.3033 & 13.3 &  44 & 64 & 17.7280 & 17.7 \\
        HD 158815 & 2016 May 03 & 240 &  8 & 20.6505 & 20.7 & 263 &  8 & 20.0293 & 20.0 \\
                  & 2017 Jun 29 & 256 &  8 & 37.3086 & 37.3 & 248 &  8 & 39.3772 & 39.4 \\
        HD 176638 & 2016 Jun 04 &  80 & 32 & 31.3344 & 31.3 &  40 & 64 & 32.9430 & 32.9 \\
                  & 2017 May 28 & 128 & 16 & 25.9954 & 26.0 &  32 & 64 & 26.8934 & 26.9 \\
        HD 182681 & 2016 May 15 & 288 &  8 & 80.0202 & 80.0 & 144 & 16 & 85.0315 & 85.0 \\
        HD 192544 & 2016 May 16 & 144 & 16 & 50.4951 & 50.5 &  72 & 32 & 53.8178 & 53.8 \\
        HD 207204 & 2016 Jun 30 &  80 & 32 & 26.5414 & 26.5 &  81 & 32 & 29.0230 & 29.0 \\
        \end{tabular}
    \caption{Observation details for each of our \sphere~sequences. In each case, the listed exposure time refers to each individual science image. Initial observations totaled $\sim$35-45 minutes of on-sky integration, and follow-up observations were often shorter depending on the magnitude and separation of the candidates identified in the initial observations.}
    \label{tab:obs}
\end{table*}

\subsection{Survey Sensitivity}
\label{sec:ccs}

We used a fake planet injection technique to calculate sensitivity as a function of radius. Fake planets were first inserted into each cleaned datacube at a variety of position angles, separations and magnitudes. We injected several fake planets into each datacube, but ensured a minimum separation of 100mas between planets so as to avoid cross contamination. For each set of fake planets, we repeated the injection with five different position angles. A PCA algorithm was applied to each of these injected datacubes as described in Section \ref{sec:dr}, and the signal to noise of each fake planet in the processed data calculated. We interpolated fake planets at the same position across magnitude, so as to find the precise magnitude that corresponds to a 5$\sigma$ detection. This process was repeated several times with up to 1/3 of the available principal components removed, so as to determine which reduction is the most sensitive to companions as a function of radius. Finally, the contrast correction term described in \citet{mawet2014} was applied to account for the small number of resolution elements at small inner working angles.

\begin{figure*}[ht]
\centering
\includegraphics[width=0.7\textwidth]{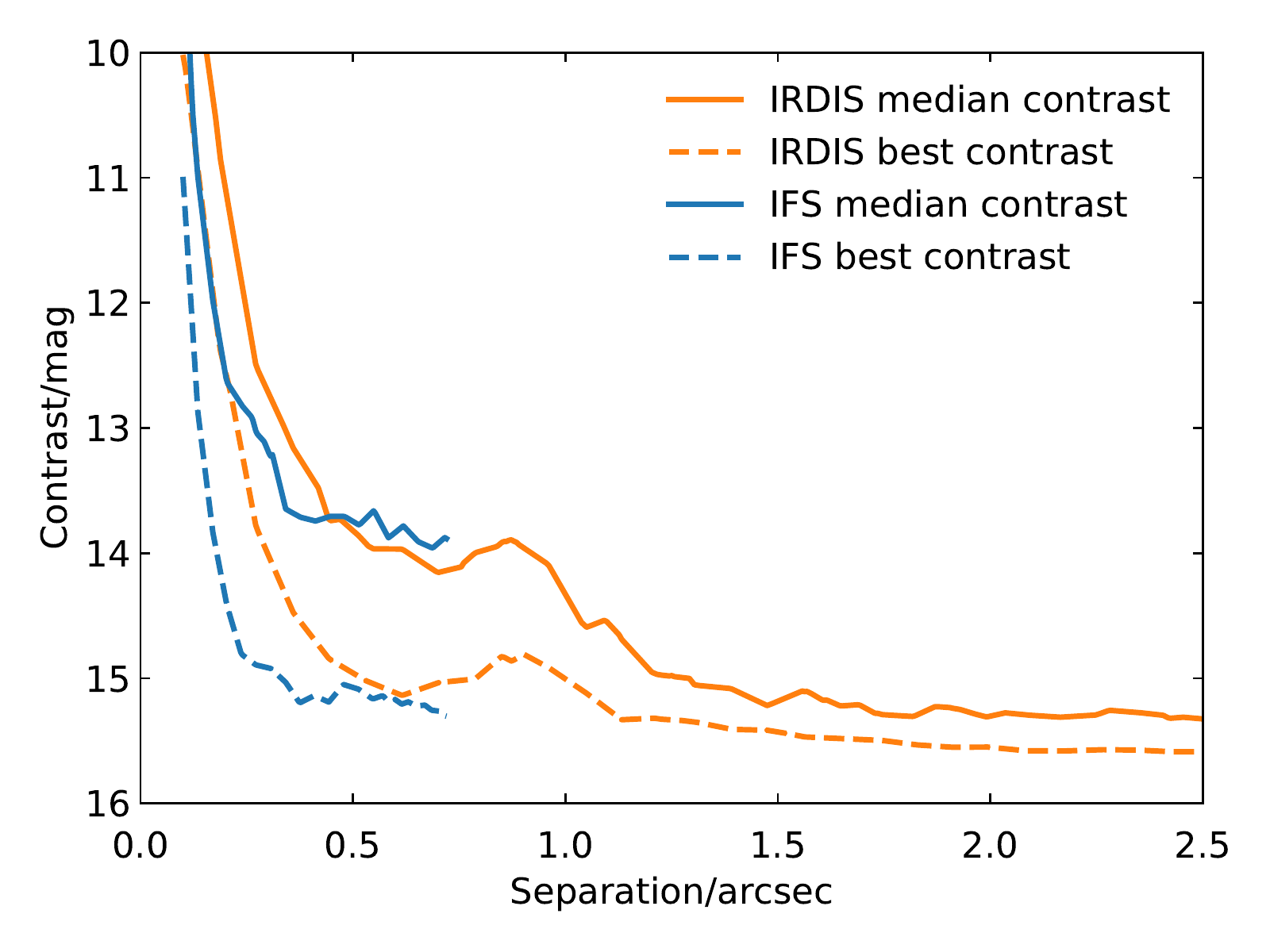}
\caption{Median and best contrasts for initial observations in this survey, with the performance of the SPHERE/IRDIS subsystem given in orange and that of SPHERE/IFS in blue. In each case, the contrast shown is the ASDI contrast, i.e. the contrast curve when both spectral and angular diversity are used to subtract speckles, and where the images are co-added across wavelength. Although the IFS slightly outperforms IRDIS in the very near-field, the instruments perform similarly beyond $\sim0.5\arcsec$. We reach typical contrasts of 13.8~magnitudes with both systems at $0.5\arcsec$}
\label{medianbestcontrast}
\end{figure*}

In this survey we reach typical contrasts of 13.8~mag at $0.5\arcsec$ with both subsystems, and 15.4~mag in the wide field at a separation of $4\arcsec$ with IRDIS. A plot of the median contrast of our survey is presented in Figure \ref{medianbestcontrast}: for this plot we only include the deepest observation of each target, which is the first epoch for the majority of targets since typically follow-up observations had shorter exposure times. The only target where the second epoch outperformed the first is HD~98363, for which the first observation was taken in poor weather conditions. Contrast curves of individual observations are available on DIVA, the Direct Imaging Virtual Archive \citep{vigan2017}, and as machine readable files provided with this article. Each file includes the measured contrast, and the 1-$\sigma$ upper and lower bounds, for regularly spaced radial separations between 0.1'' and the edge of the field of view.

\begin{figure*}[ht]
\centering
\includegraphics[width=0.9\textwidth]{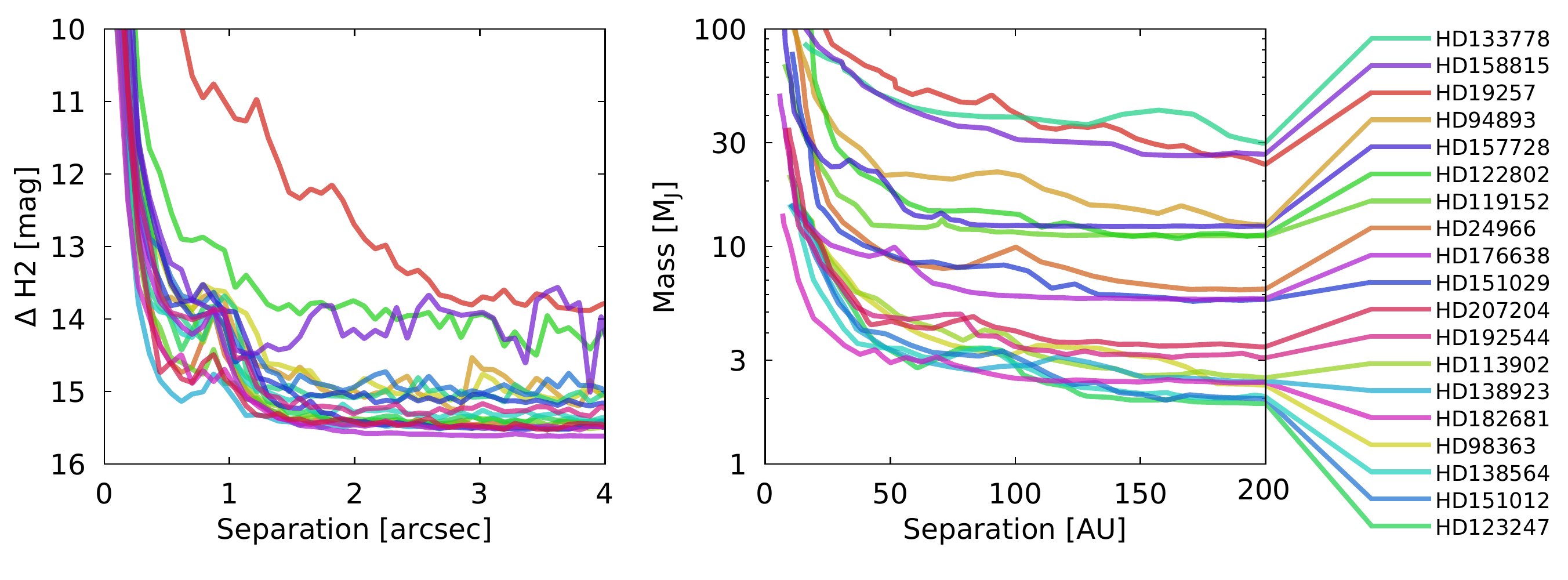}
\caption{\textit{Left:} Contrasts for each individual observation. The contrast for HD19257 is markedly less than for the other observations, due to the presence of a bright companion 1.3'' from the star. \textit{Right:} Mass sensitivity of the survey, as a function of projected separation. For this conversion we use the COND models \citep{baraffe2003} for objects below 1700K and the DUSTY models \citep{chabrier2000} otherwise. Contrasts are only shown out to 4'' and 200au since sensitivity is largely flat beyond these separations; our field or view extends to $\sim$6'' and between $\sim$250au and $\sim$850au depending on the stellar distance. The median survey sensitivity is 10.4\Mjup~at 25au and 5.9\Mjup~at 100au}
\label{masscontrast}
\end{figure*}

SPHERE/IRDIS contrast limits are converted to approximate mass limits using the COND models \citep{baraffe2003} for temperatures below 1700K and the DUSTY models \citep{chabrier2000} otherwise. This represents a somewhat optimistic scenario where planets formed following a `hot-start' process, with high initial entropy and effective temperature at formation. For this conversion we also use the age determined for each target, as discussed below. Since no age could be confirmed for HD18378, we did not calculate a mass sensitivity for this target. Contrasts and mass limits for each individual target except HD18378 are shown in Figure \ref{masscontrast}. For HD19257 the contrast is reduced due to the presence of a very bright companion close to the target star (see Section \ref{sec:19257}). We are least sensitive to low mass companions for HD158815 and HD133778, which are the oldest two targets in the survey, and for HD19257, which is also relatively old and has a bright companion reducing the contrast sensitivity. The 7 targets with the best mass sensitivity are all members of nearby moving groups, and are therefore all very young targets. The median survey sensitivity is 10.4\Mjup~at 25au and 5.9\Mjup~at 100au.

\section{Methods \& Results}
\label{sec:methods}

\subsection{Age Determination}
\label{sec:ages}

Determining accurate ages is vital in high contrast imaging, since the relationship between brown dwarf/planet mass and luminosity is dependent on the age of the object. The target age therefore plays a role in determining both the contrast limits that the survey reaches, and in the case of low mass companions, the properties of those companions. Targets are not selected explicitly for youth, as described above, but we nonetheless expected a bias towards young stars since the occurrence of extrasolar dust is much higher for young stars \citet{kennedywyatt2012}.

\begin{figure*}[ht]
\centering
\includegraphics[width=0.95\textwidth]{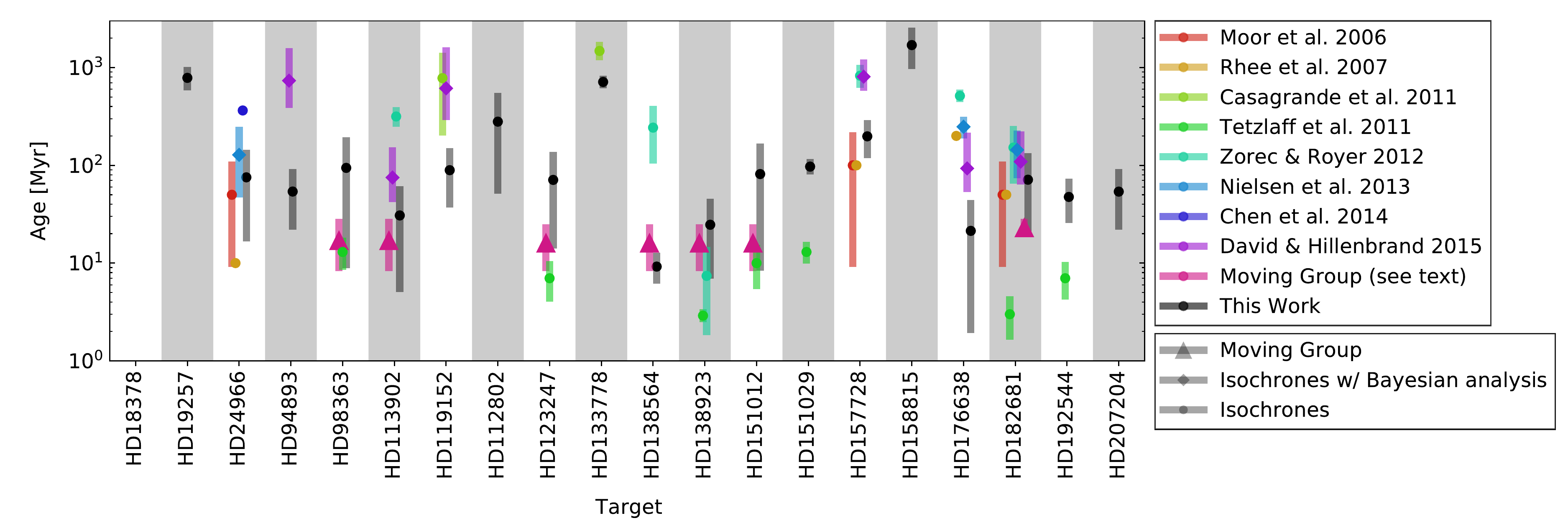}
\caption{Literature ages for candidates in this survey, from a variety of sources. In most but not all cases, the literature age determinations are consistent. We also independently generate ages for each target with the \texttt{isochrones} package \citep{mortonisochrones} using BVGJHK photometry and \gaia~parallaxes. These ages broadly consistent with literature values, and are only used for targets that are not moving group members and that have one or fewer literature age references. The age used in analysis for each object is listed in Table \ref{tab:dustytargets}. References are \citet{moor2006,rhee2007,casagrande2011,tetzlaff2011,zorecroyer2012,nielsen2013,chen2014,davidhillenbrand2015}.}
\label{fig:allages}
\end{figure*}

To determine ages for the sample, we first turned to the literature. The targets in our survey have anywhere between 0 and 8 independent age determinations in the literature, which are usually but not always consistent for the same target, as demonstrated in Figure \ref{fig:allages}. These literature ages are calculated by two classes of methods. First, several of the targets in this work are members of moving groups, with well constrained ages. Many of the targets have previously been identified as members of the Scorpius-Centaurus association, which has a well defined and well studied age spread \citep[see e.g.][]{pecautmamajek2016}. We also searched for previously unidentified moving group members using the BANYAN-$\Sigma$ tool \citep{gagne2018}, which compares galactic position and space velocity to known nearby, young associations. We found that HD182681 has a 90.2\% likelihood of being a member of the $\beta$-Pictoris moving group. A number of other works also quote young ages for this target (see Figure \ref{fig:allages}), confirming that this object is indeed a moving group member and not aligned with the group by chance. 

Secondly, many of the targets have had ages determined by comparison to isochrones. However, this approach can lead to biased ages, since it does not account for the non-uniform distribution of masses or the fact that time is non-linear on an HR diagram \citep{pont2004,takeda2007}. Both \citet{nielsen2013} and \citet{davidhillenbrand2015} therefore use a more sophisticated Bayesian inference technique to extract age designations from isochrones. In some cases no isochronal ages have been previously determined, and so we fit ages using the \texttt{isochrones} tool \citep{mortonisochrones}, using the \gaia~parallax and BVGJHK photometry from Tycho, 2MASS and \gaia~\citep{hog2000,skrutskie2006,gaia_dr1a}. To verify these values, we repeat the fits using only the BVG, and then only the JHK photometry, and check for consistency.

In Table \ref{tab:dustytargets}, we report our final choice of age for each target, and a quality flag with value between 1 and 3, indicating how reliable this age is, with 3 being the most reliable. We assign ages by the following procedure:

\begin{itemize}
\item If the object is in a moving group, we use the age of that moving group. These objects are assigned an age quality flag 3.
\item If a Bayesian isochronal age exists in the literature, we use this age designation. We use the ages from \citet{davidhillenbrand2015} where possible, and the ages from \citet{nielsen2013} otherwise. These objects are assigned an age quality flag 2.
\item For the remaining targets, there are at most one literature age reference, and we determine independent ages for each of these objects using \texttt{isochrones}. We use the \citet{casagrande2011} age for HD133778, which is discussed further below, and our own ages for the other targets. These objects are assigned an age quality flag 1. We show our estimated \texttt{isochrones} ages for all of the targets in this work in Figure \ref{fig:allages} to demonstrate their reliability. The majority of our age estimates are consistent with other literature estimates for these targets. Of the seven targets in moving groups, our estimates are consistent to within 1$\sigma$ of the moving group age for five targets, and consistent to within 1.5$\sigma$ for the remaining two.
\item HD18378 is an evolved star, and so it is not possible to extract a useful age determination using \texttt{isochrones} (see Section \ref{sec:18378} for a detailed discussion). This is the most evolved star in our sample, and in this case the excess infrared flux might be due to mass loss rather than the presence of a debris disk.
\end{itemize}

The age analysis for HD~133778 warrants special discussion, both since there is contradictory literature and since the object has a co-moving companion (see Section \ref{sec:133778}). The star is listed as a Sco-Cen member in \citet{degeus1990}, but not in the more recent works such as \citet{dezeeuw1999,rizzuto2011}. \citet{casagrande2011} find the star to have an age $\sim$1.5~Gyr and a mass $\sim1.73\Msun$ using isochronal analysis. Since the target RV ($-37\kmpers$, \citealt{nordstrom2004}, \citealt{torres2006}) is inconsistent with the values expected for a Sco-Cen member (typically +5\kmpers, \citealt{debruijne1999}) and the star is lithium poor \citep{torres2006}, we use the older age as listed in \citet{casagrande2011}. This suggests that the object was a late A type star, but has recently left the main sequence.

\subsection{Candidate Companions}

We identified candidate companions by visually inspecting each PCA-processed image. To ensure an optimum balance between removal of stellar signal and preservation of companion signal, we visually inspected data products with 1, 6 and 20 principal components removed. Once candidates are flagged, relative astrometry and photometry can be extracted from the IRDIS images. To determine the relative astrometry, we measured the companion positions by fitting Gaussians to the sources. The stellar position behind the coronagraph is inferred using the star center calibration frames. We assume that the candidate companion positions can be determined to an accuracy of 0.1pix, and that the position of the star behind the coronagraph can be determined to 0.2pix by using the star center calibration frames. This is in line with other SPHERE analyses; in particular \citet{maire2016} determined that the star position is accurate to better than 0.1pix ($\sim$1.2mas). We converted detector positions to astrophysical separations following the ESO SPHERE User Manual\footnote{https://www.eso.org/sci/facilities/paranal/instruments/sphere/doc.html}: we used a platescale of 12.255$\pm$0.021mas/pix and a true north correction of -1.700$\pm$0.076$^\circ$ for data before 2015 December, and -1.75$\pm$0.08$^\circ$ for data after 2016 February. The uncertainties from the point-finding and from the detector properties are similar in magnitude, with the uncertainty on the point-finding dominant for close candidates, and the uncertainty on detector properties dominant for more distant sources.

We determined relative photometry of the candidate companions by measuring the flux within a small aperture centered on the candidate positions, and comparing this value to the measured brightness of the host stars, measured using the same aperture, in the flux calibration frames. We accounted for self-subtraction by injecting several fake planets into the data at varying separations and known magnitudes. These were then extracted using the same aperture photometry procedure, and we used the measured magnitude to calculate a self-subtraction fraction as a function of radius, which can be applied to the real candidate companions. 

For HD19257, the candidate companion (later confirmed to be co-moving, see Section \ref{sec:19257}) is sufficiently bright that it is clearly visible in the flux calibration frames, and the signal from the companion is severely distorted in the PCA processed images due to extensive self-subtraction. In this case we instead measured the photometry and astrometry directly from the cleaned and pre-processed flux calibration frames to avoid biases and self-subtraction. For the HD18378B and HD133778B companions, which are also confirmed to be co-moving (see Sections \ref{sec:18378} and \ref{sec:133778}, we refine the photometry by performing negative fake planet subtraction and iterating with a Levenberg-Marquardt least-squares minimization. This process is identical to that used for the IFS spectral extraction, as described in Section \ref{sec:ifsspectra}. This procedure shows results in good agreement with the simpler approach (i.e.~aperture photometry with a correction to account for self-subtraction) and is significantly more computationally expensive, and so is only used for these two co-moving companions and not for the unconfirmed candidate companions or background stars.

In this survey we identified a total of 162 candidates, with 98 of these candidates associated with the target HD~158815 which is very close to the galactic plane (galactic latitude \textit{b}=+6.7$^\circ$). The median number of candidate companions per target in this survey is 1, while the mean is 8. Of the 162 candidates identified, 3 are confirmed to be co-moving companions in the stellar mass regime (namely HD~18378~B, HD~19257~B and HD~133778~B; see Sections \ref{sec:133778}-\ref{sec:18378}).

\subsubsection{Common Proper Motion Analysis}

\begin{figure*}[ht]
\centering
\includegraphics[width=\textwidth]{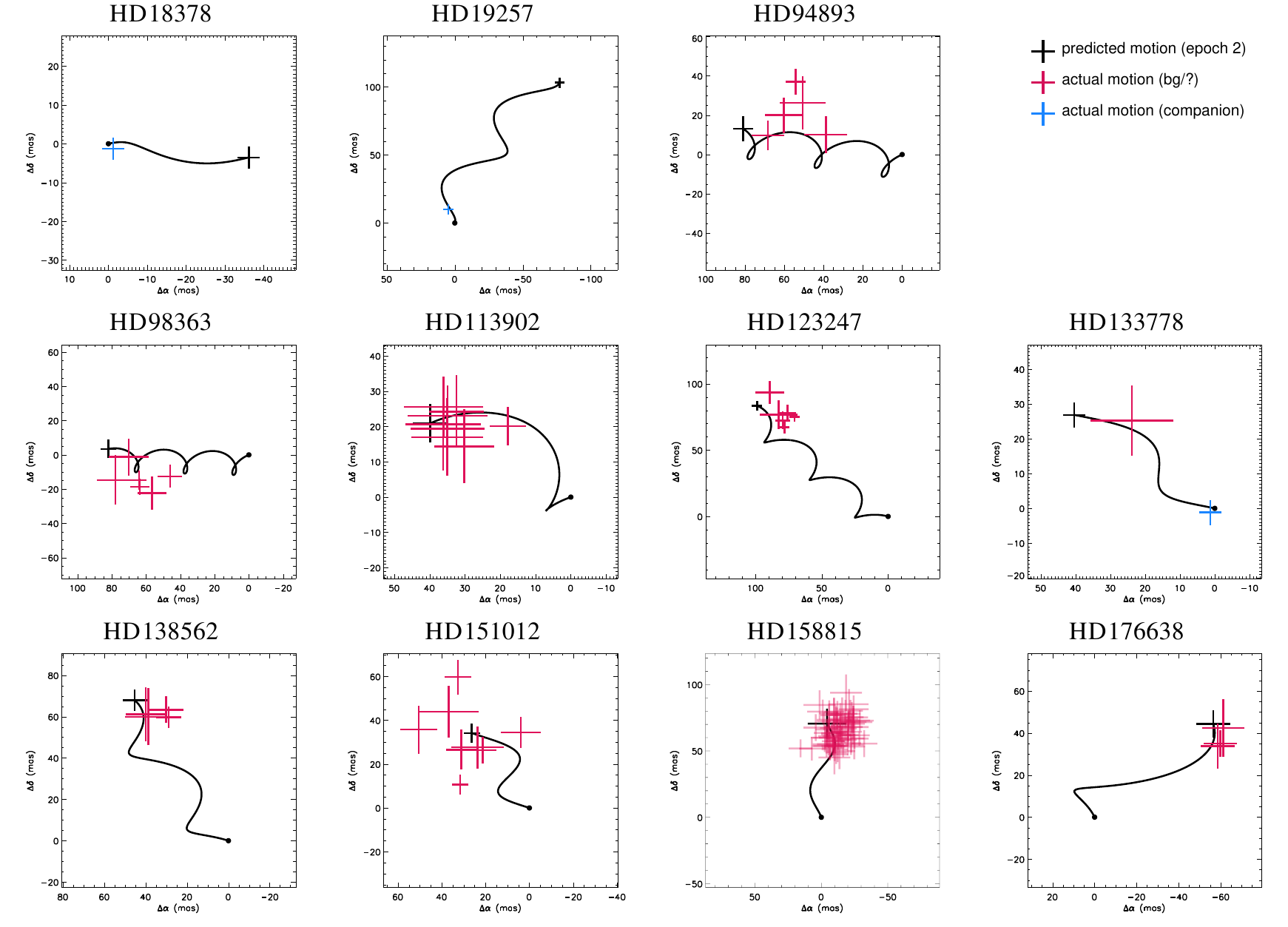}
\caption{Combined common proper motion plots for those companions for which we have two astrometric measurements. In each case the epoch 1 positions are aligned to [0,0] and plotted with a black circle, and the change in relative astrometry (i.e., the vector motion) between the epochs is indicated in red for background objects and blue for companions. The black line and black cross indicate the expected motion of a background star, and the expected position of background stars at the second epoch, with the dominant error being the accuracy of our astrometry. Note that this position is the same for all background objects, since it is derived from the proper motion and parallax of the foreground object. The HD158815 points are partially transparent for clarity. Some systematic errors are seen, which are discussed in the text.}
\label{fig:cpm_all}
\end{figure*}

In Figure \ref{fig:cpm_all} we present combined common proper motion plots for each target where we detect candidates. Host star proper motions are taken from the Gaia DR2 catalog \citep{gaia_dr2}. In each case we plot a point indicating the final astrometric position of each individual candidate relative to its initial position, i.e., we show the vector position change of the companion between epochs. We also show the predicted path and final position for a stationary background object relative to the target star. Candidates are color-coded to indicate their designation as companion or background (but note that the 31 candidates with only one astrometric measurement are not indicated in this plot). In most cases, the final positions of candidates are clustered around the predicted position for a background object, and we conclude that these objects are not associated with the target star. 

For a few of the sources, the astrometry does not show strong agreement with the background hypothesis, either because there is significant scatter between the candidate companions (such as HD151012), or because the median candidate companion motion vector is offset from the predicted motion vector (such as HD94893). For HD151012, this is likely an issue in the centering of individual point sources: many of the points or only marginally above the detection limit, and are redetected at below a conventional 5$\sigma$ detection threshold in the second epoch, where the sensitivity is marginally lower than the first epoch. Indeed, the four points that are most divergent from the background hypothesis are also the four faintest points that are redetected. This indicates that the 0.1px accuracy in location is too optimistic for the faintest of point sources targets. For HD94893, the median candidate vector is offset by 28mas, or slightly over 2 pixels, from the background hypothesis. A similar effect is seen for HD98363 and HD123247, where the offsets are 26mas and 21mas. The average position error for these sources are 13mas, 11mas and 9 mas respectively. This indicates that we are seeing a systematic error at the 2$\sigma$ level in each case. The source of this systematic error is unclear. It is likely associated with either the stellar position drifting behind the coronagraph, or a small misalignment of the north angle. The median candidate vector is offset from the common proper motion hypothesis by 58mas, 110mas and 65mas for the three targets, and so in each case the background hypothesis remains a much better fit than the common proper motion hypothesis.

\begin{figure*}[ht]
\centering
\includegraphics[width=\textwidth]{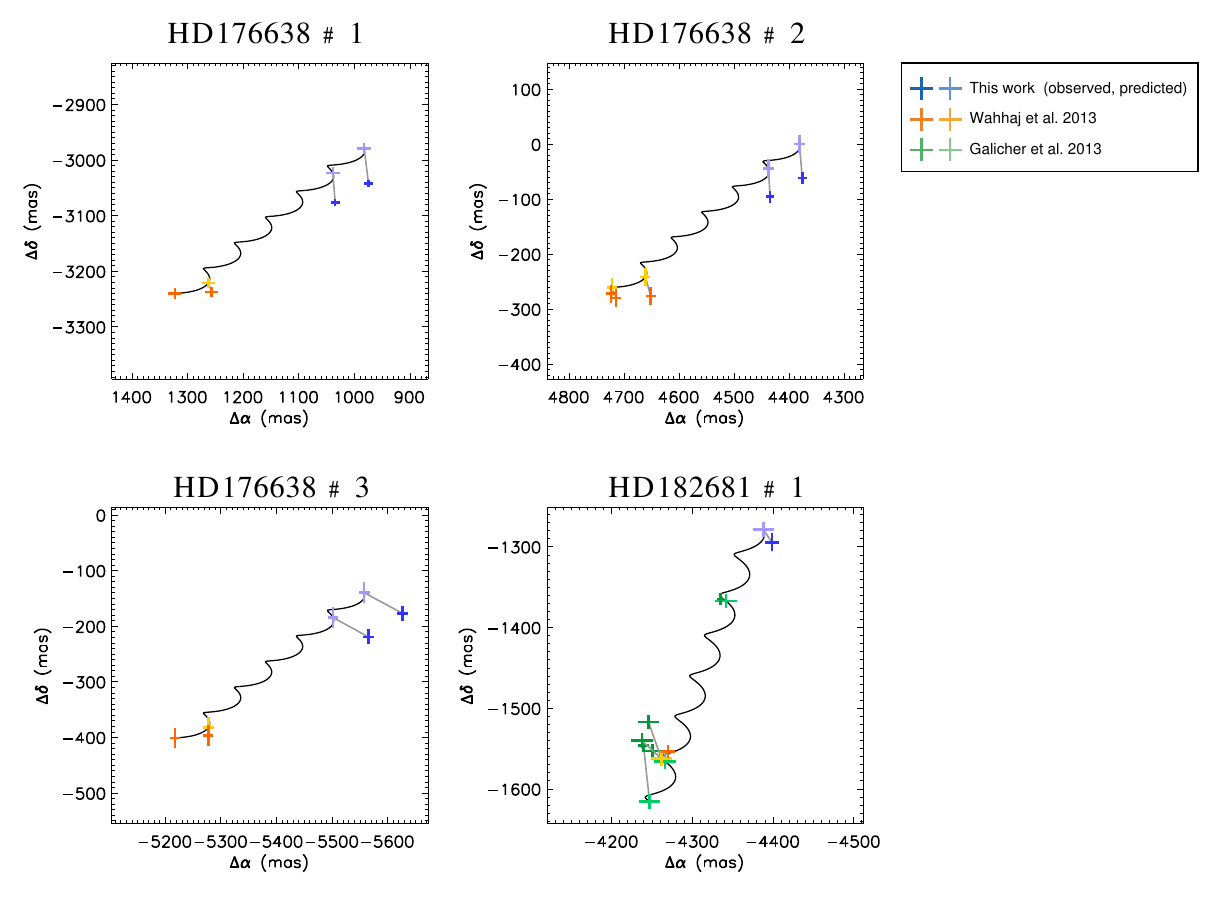}
\caption{Common proper motion plots for those candidates with archival astrometry. The black line indicates the predicted path for a stationary background object, which is measured relative to the first archival epoch for HD176638, and relative to the second epoch for HD182681 since the first epoch is inconsistent with other early measurements. In each case the orange, green and blue points indicate astrometry from \citet{wahhaj2013}, \citet{galicher2016} and this work respectively, with darker points being measurements and lighter points being the predicted background position at each epoch. For HD176638 we see a small offset between the astrometry reported in this work and that from \citet{wahhaj2013}, common in comparative astrometry between different instruments with their own systematics. For HD182681 we see broadly good agreement between all epochs, albeit with significant scatter between the earlier measurements. Our astrometry is consistent with previous works, and with each of these candidates being a background source.}
\label{archivalcpm}
\end{figure*}

Archival high resolution data exists for HD24966, HD157728, HD176638 and HD182681 \citep{nielsen2013, wahhaj2013, galicher2016}, with the first two publications reporting on the same data. For both HD24966 and HD157728, we did not detect any candidate companions, a result in agreement with \citet{wahhaj2013}, who report no candidates for HD24966 and a single wide candidate to HD157728 at 11.4'', well outside our field of view. For HD176638 and HD182681, all of the archival candidates in our field of view have previously been reported as backgrounds, and our astrometry shows good agreement with the archival measurements, as demonstrated in Figure \ref{archivalcpm}. We see a small offset between our measurements of the three HD176638 candidate companions and the stationary background object hypothesis: such small systematic errors are common when comparing different instruments with their own systematics. For HD182681, our measurements show good agreement with the archival astrometry.

Based on multi-epoch astrometry and comparisons to archival astrometry, we are able to confirm 119 of the candidate companions as background sources.

\subsubsection{Color-Magnitude Analysis}
\label{sec:cmd}

\begin{figure*}[ht]
\centering
\includegraphics[width=7cm]{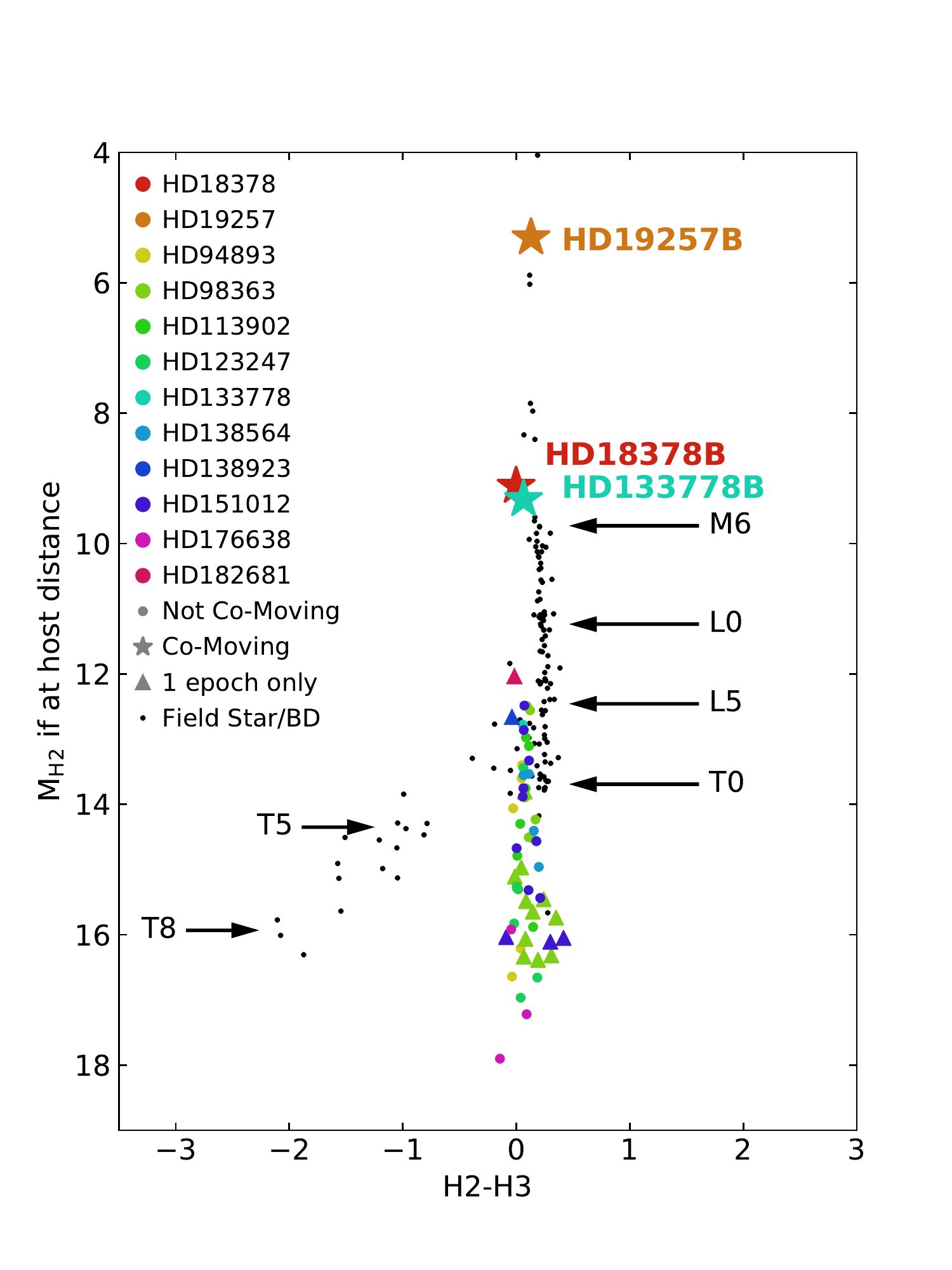}
\includegraphics[width=7cm]{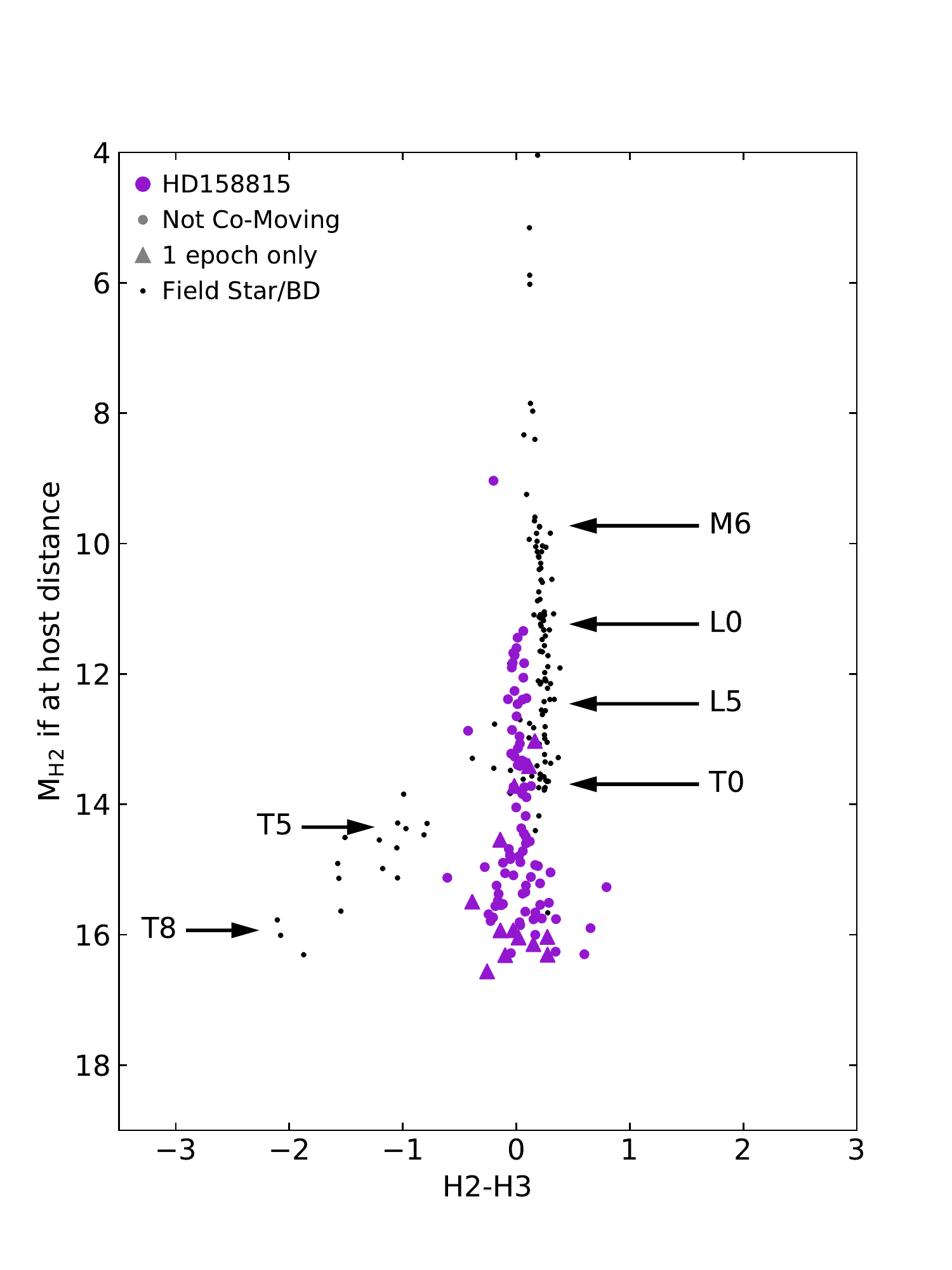}
\caption{Color-Magnitude Diagrams for candidate companions in this survey. For each candidate we plot the absolute magnitude if the object were at the distance of the host star against the H2-H3 color. Black points indicate the expected trend for late type stars: objects of spectral type T0 or later start to have significant methane absorption, and a correspondingly negative H2-H3 color. Faint candidates with H2-H3$\sim$0 are inconsistent with being at the distance of the host star, and are therefore presumed to be background objects rather than co-moving companions. Specifically, we assume all candidates with M$_\textrm{H2} >$ 15 and H2-H3 $>$ (15-M$_\textrm{H2}$)/1.5 are background objects. Those candidates without common-proper motion data, for which the color-magnitude analysis is particularly important, are indicated with triangular points. The three large stars indicate the location of the three co-moving companions on the CMD, which are discussed further in Sections \ref{sec:18378}-\ref{sec:19257}.}
\label{fig:cmd}
\end{figure*}

Our second approach to differentiating companions and background stars is based on the colors of candidates. We calculated the H2-H3 color of each candidate, and the absolute magnitude in H2 under the assumption that the candidate is at the distance of the host star. The SPHERE H3 filter lies on a methane absorption band, and so for sufficiently cold objects (spectral type T or later) we expect to see a negative H2-H3 color, while earlier type objects have a color of $\sim$0. We can therefore use this color to differentiate faint background stars from faint companions, since for faint objects a color $\sim$0 indicates that the object is not at the distance of the host, and is instead a background object of earlier spectral type, with no methane absorption. In Figure \ref{fig:cmd}, we show the CMD for candidates in our survey. We show the large number of candidate companions to HD158815 in a separate panel for clarity. Those objects without CPM data, i.e. those for which the color-magnitude analysis is particularly important, are highlighted as triangles in the figure. For reference, we also plot CMD positions of late type stars and field brown dwarfs from the SpeX Prism Library\footnote{ \url{http://pono.ucsd.edu/~adam/browndwarfs/spexprism}}. We selected all available objects with high resolution spectral data, and integrated the flux of each object across the SPHERE H2 and H3 filters. Objects of spectral type M and L have color $\sim$0 in these filters, and there is a sharp turn-off at L-T transition with T objects showing negative H2-H3 colors due to methane absorption in the atmosphere. Based on the positions of the SpeX objects on this CMD, we assume all candidates in our survey with M$_\textrm{H2} >$ 15 and H2-H3 $>$ (15-M$_\textrm{H2}$)/1.5 are background objects. Finally, the three confirmed co-moving companions detected in this survey are highlighted with stars in the plot: based on H2 magnitude alone, HD~133778~B and HD~18378~B have spectral type $\sim$M4, while HD~19257~B is $\sim$M0. These companions and their properties are discussed further in Sections \ref{sec:133778}-\ref{sec:18378}.

Using this CMD approach, we are able to confirm an additional 16 candidate companions as background stars.

\subsubsection{Unconfirmed Candidate Companions}
\label{sec:unconfirmedcand}

Using the combination of astrometric and color-magnitude analysis, we confirm 144 of the 162 candidate companions as background objects. Three candidates are confirmed to be co-moving companions. The remaining 15 objects are likely background stars since all are widely separated from their host stars, but more data would be required to measure astrometry relative to the host and thus confirm this conclusion. Twelve of these unconfirmed objects are candidate companions to HD158815, and are highly likely to be backgrounds, due to the crowded background field of this target. The closest separation of an unconfirmed candidate companion to its host star is 2.1'', and the unconfirmed candidates follow the approximately radial distribution that would be expected for background stars. In Appendix \ref{appendix:targetnotes} we include additional notes for each individual target explaining the final designations of each candidate as a companion, a background object or a likely background object. Individual candidate astrometry is available in Table \ref{tab:astrometry} and in the online DIVA archive \citep{vigan2017}.

\begin{table*}[ht]
    \centering
    \footnotesize

    \begin{tabular}{llllllllllll}
		Target & Date & N & PA[deg] & PA\_err & Sep[mas] & Sep\_err & RA[mas] & RA\_err & Dec[mas] & Dec\_err & Status \\
		\hline \hline
		HD18378 & 2016-09-15 & 1 & 279.35 & 0.40 & 418.2 & 2.8 & -412.6 & 2.8 & 68.0 & 2.9 & C \\
		HD18378 & 2017-08-29 & 1 & 279.15 & 0.40 & 419.2 & 2.8 & -413.8 & 2.8 & 66.7 & 2.9 & \\
		HD19257 & 2015-08-26 & 1 & 127.70 & 0.16 & 1292.9 & 2.9 & 1022.9 & 3.2 & -790.7 & 3.3 & C \\
		HD19257 & 2017-09-01 & 1 & 127.28 & 0.16 & 1291.0 & 2.9 & 1027.2 & 3.2 & -782.0 & 3.3 & \\
		HIP94893 & 2015-04-11 & 1 & 275.69 & 0.15 & 2547.5 & 5.2 & -2534.9 & 5.2 & 252.5 & 6.5 & BG\_cpm \\
		HIP94893 & 2018-01-07 & 1 & 276.66 & 0.15 & 2497.6 & 5.1 & -2480.8 & 5.1 & 289.6 & 6.4 & \\
		HIP94893 & 2015-04-11 & 2 & 218.04 & 0.14 & 3647.9 & 6.9 & -2247.6 & 8.2 & -2873.1 & 7.7 & BG\_cpm \\
		HIP94893 & 2018-01-07 & 2 & 217.27 & 0.14 & 3598.4 & 6.8 & -2179.3 & 8.2 & -2863.4 & 7.7 & \\
    \end{tabular}
    \caption{Astrometry for each individual candidate at each epoch. The full table is available in machine readable form. In the status column, we indicate companions with \textit{C}, candidates confirmed as backgrounds based on their proper motion as \textit{BG\_cpm}, those confirmed as backgrounds based on their CMD position as \textit{BG\_color} and finally those candidates that are unconfirmed as \textit{?} (although note that these are all likely to be background stars as detailed in Section \ref{sec:unconfirmedcand}).
}
    \label{tab:astrometry}
\end{table*}

\subsection{IFS spectral extraction}
\label{sec:ifsspectra}

For companions within the IFS field of view, we perform a dedicated analysis to extract a spectrum that can be compared to empirical or theoretical templates. To avoid biases usually induced by spectral differential imaging \citep{racine1999} in high-contrast spectroimaging or spectroscopic data \citep[e.g.][]{thatte2007,vigan2012scr}, we perform an ADI analysis channel-by-channel, that is considering each wavelength in the IFS data independently. For the speckle subtraction we use a full-frame PCA with typically only a few modes subtracted.

At each wavelength, the precise astrometry and photometry of the candidate companions are estimated using negative fake planet subtraction in the pre-processed data cube \citep[e.g.][]{marois2010b}. A rough estimation of the object position and contrast is first performed using a 2D Gaussian fit. These initial guesses are then used as a starting point for a Levenberg-Marquardt least-squares minimization, where the position and contrast of the negative fake companion are varied to minimize the residual noise after ADI processing in a circular aperture of radius $\lambda/D$ that is centered on the position of the companion. When a minimum is reached, the position and contrast of the fake companion are taken as the optimal values for the astrometry and photometry. This approach has already been used reliably for various companions over a wide range of contrast \citep{bonnefoy2011,zurlo2016,vigan2016}. 

To estimate the error bars on the astrophotometry, we inject fake positive companions at the same separation but different position angles than the real candidate, with the optimal photometry. Then, the same minimization procedure as described previously is applied to these fake positive companions to derive their best fit photometry and astrometry. This operation is repeated at 100 different position angles, each time with only a single positive fake companion injected to avoid any bias in the results due to the presence of multiple companions at the same separation. The error bars on the astrometry and photometry of the real candidate are then estimated using the 1$\sigma$ deviations measured on the determination of the astrophotometry of the positive fake companions. Although this procedure cannot capture the uncertainties related to the variation of the observing conditions, it enables a reliable estimation of the uncertainties related to the image processing and signal extraction.

Only the photometry from this spectral extraction process is used in our analysis, although we note that the astrometry is consistent with that derived from the IRDIS data.


\section{Confirmed Companions}
\label{sec:companions}

Three stellar companions are identified with \sphere~in this survey, and we find evidence in the \gaia~catalog of three additional very-wide stellar companions, which are outside of the field of view of the SPHERE instrument. Table \ref{tab:companions} presents an overview of the systems with stellar companions, and we also discuss these companions individually below.

\begin{table*}[ht]
    \centering
    \footnotesize

    \begin{tabular}{l|ccc|ccc}
    	 & HD18378 & HD19257 & HD133778 & HD98363 & HD138564 & HD151012 \\
    	\hline \hline
    	Detected by & SPHERE & SPHERE+Gaia & SPHERE & Gaia & Gaia & Gaia\\
    	Host SpT & K2III & A5V & G2V & A2V & B9V & B9.5V \\
    	Companion SpT & M4$\pm$2 & early M? & M4$\pm$1 & --- & --- & --- \\
    	T$_\textrm{eff}$ Comp [K] & --- & n/a & --- & 4360 & 3590 & 4100 \\
    	\hline

    	\textbf{SPHERE Epoch 1} & \textbf{2016-09-15} & \textbf{2015-08-26} & \textbf{2016-05-16} & \\

    	Separation [arcsec] & 0.418$\pm$0.003 & 1.293$\pm$0.003 & 0.989$\pm$0.003 & \\
    	Projected Sep [au] & 218$\pm$3 & 96.2$\pm$0.5 & 162$\pm$2 & \\

    	Position Angle [$^\circ$] & 279.4$\pm$0.4 & 127.70$\pm$0.16 & 88.3$\pm$0.2 & \\

    	\hline

    	\textbf{SPHERE Epoch 2} & \textbf{2017-08-29} & \textbf{2017-09-01} & \textbf{2017-06-22} & \\

    	Separation [arcsec] & 0.418$\pm$0.003 & 1.291$\pm$0.003 & 0.990$\pm$0.003 & \\
    	Projected Sep [au] & 218$\pm$3 & 96.1$\pm$0.5 & 162$\pm$2 & \\
    	Position Angle [$^\circ$] & 279.2$\pm$0.4 & 127.28$\pm$0.16 & 88.4$\pm$0.2 & \\

    	\hline

    	\textbf{Gaia Epoch} &  & \textbf{Gaia DR2} &  & \textbf{Gaia DR2} & \textbf{Gaia DR2} & \textbf{Gaia DR2} \\

    	Separation [arcsec] &  & 1.2537$\pm$0.0012 &  & 49.65 & 22.59 & 34.09 \\
    	Projected Sep [au] &  & 93.1$\pm$0.4 &  & 6860$\pm$70 & 2312$\pm$15 & 3670$\pm$20 \\
    	Position Angle [$^\circ$] &  &130.6 &  & 304.0 & 49.6 & 144.6 \\

    	\hline

    	$\Delta$H2 [mag] & 10.76$\pm$0.05 & 3.28$\pm$0.04 & 8.16$\pm$0.03 & \\
    	$\Delta$H3 [mag] & 10.61$\pm$0.05 & 3.16$\pm$0.04 & 8.06$\pm$0.03 & \\

    	$\Delta$G [mag] &  & 4.53$\pm$0.03 &  & 3.625$\pm$0.002 & 6.7315$\pm$0.0005 & 7.1179$\pm$0.0006 \\

		\hline
    	Est. Orbital Period [yr]\textsuperscript{a} & 2600 & 630 & 1400 & 350,000 & 73,000 & 150,000 \\
    	Est. Orbital Motion [mas/yr] \textsuperscript{a} & 1 & 13 & 4 & 1 & 2 & 1 \\
    	\vspace{0.3cm}
    \end{tabular}

    \caption{Summary of the stellar companions to stars in this survey. The companion to HD98363 is commonly known as Wray-15 788; companions to HD19257, HD138564 and HD151012 have only Gaia names and are Gaia~DR2 123185851097644416, Gaia~DR2 6003218283764194304 and Gaia~DR2 6034203483517407616 respectively. The quoted H-band contrast is in the SPHERE H2 filter for SPHERE targets, and is not listed for those objects we do not detect with SPHERE; the $\Delta$G values are the contrast in the main \gaia~filter. T$_\textrm{eff}$ values are the \gaia~temperature fits, where available. Full epoch astrometry, including the relevant dates, are also listed for both the SPHERE observations and the \gaia~DR2 measurements. \\ \textbf{a:} the estimated orbital period and motion values rely on many assumptions, most notably that the projected separation is the real separation, and the orbit is circular and the projected. These values are provided only to give a sense of the parameter space of these companions.}
    \label{tab:companions}
\end{table*}

\subsection{HD~133778~B: a mid-M companion to an star that recently left the main sequence}
\label{sec:133778}

\begin{figure*}[ht]
\centering
\includegraphics[width=0.6\textwidth, trim=1.7cm 0.8cm 1.5cm 0.4cm, clip]{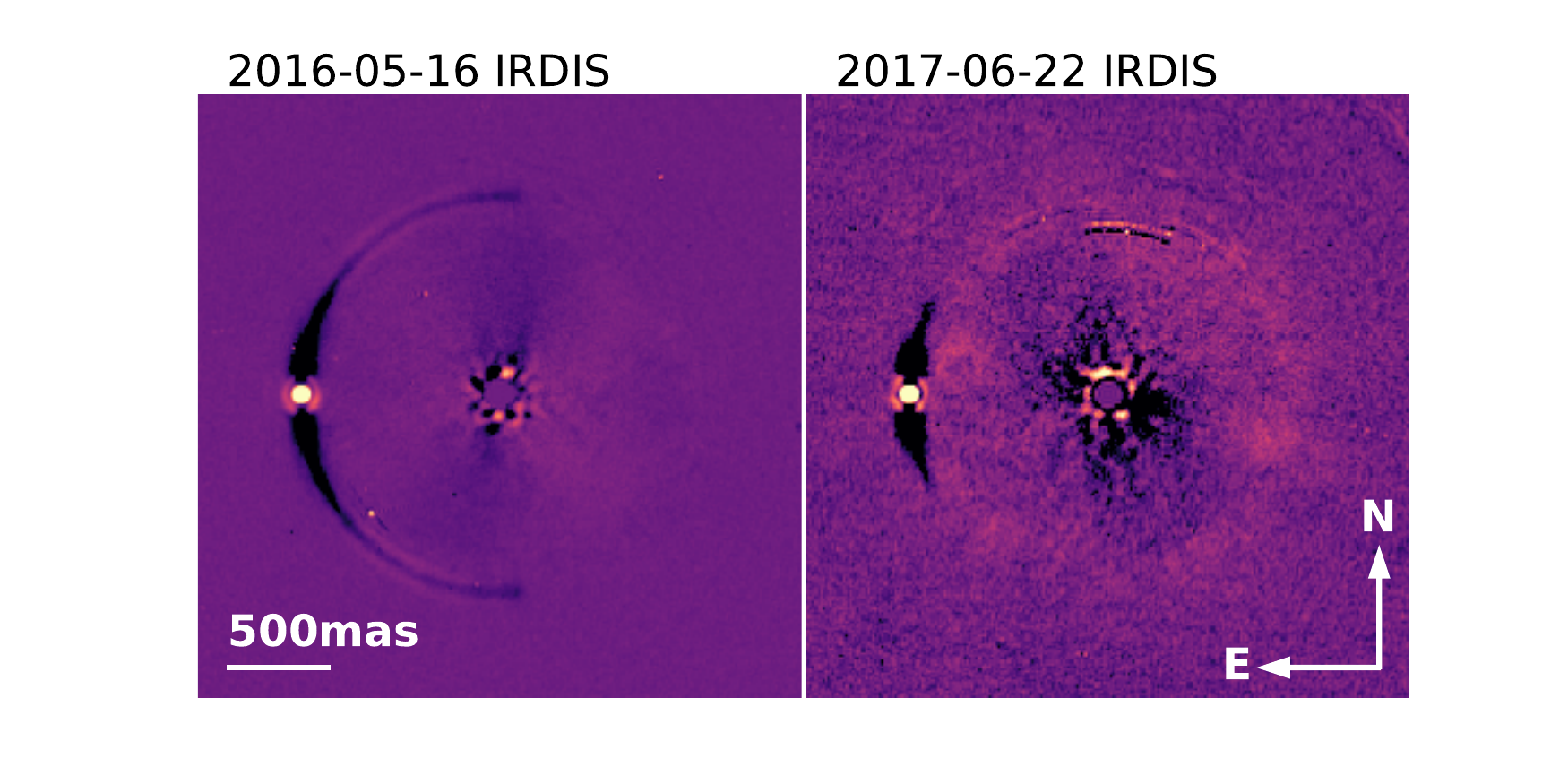}
\caption{Discovery images of HD~133778B. The left and right panels show data from 2016-05-16 and 2017-06-22 respectively. In each case, the SPHERE/IRDIS data is reduced using a PCA algorithm to remove stellar speckles, and the H2 and H3 data are coadded. The host star is in the center behind the coronagraphic mask, and the companion is clearly visible to the left hand side of the image. Note that the bright circular ring at $\sim$0.9'' is the ring of speckle noise at the edge of the AO control radius, smeared due to the rotation between images, and is not related to the debris disk in this system.}
\label{fig:hd133778}
\end{figure*}

We identify a companion to HD~133778 for the first time (Figure \ref{fig:hd133778}), which shows clear evidence for common proper motion with the host (see Figure \ref{fig:cpm_all} above). The companion is at a projected separation of 989$\pm$3mas (=162$\pm$2au at the host distance) and a position angle of 88.3$\pm$0.2$^\circ$, and we measure a movement of 2.7mas relative to the host between our two datasets, a value smaller than the astrometric precision of SPHERE. As such, we do not attempt to fit an orbit at this point.

\begin{figure}[ht]
\centering
\includegraphics[width=0.49\textwidth]{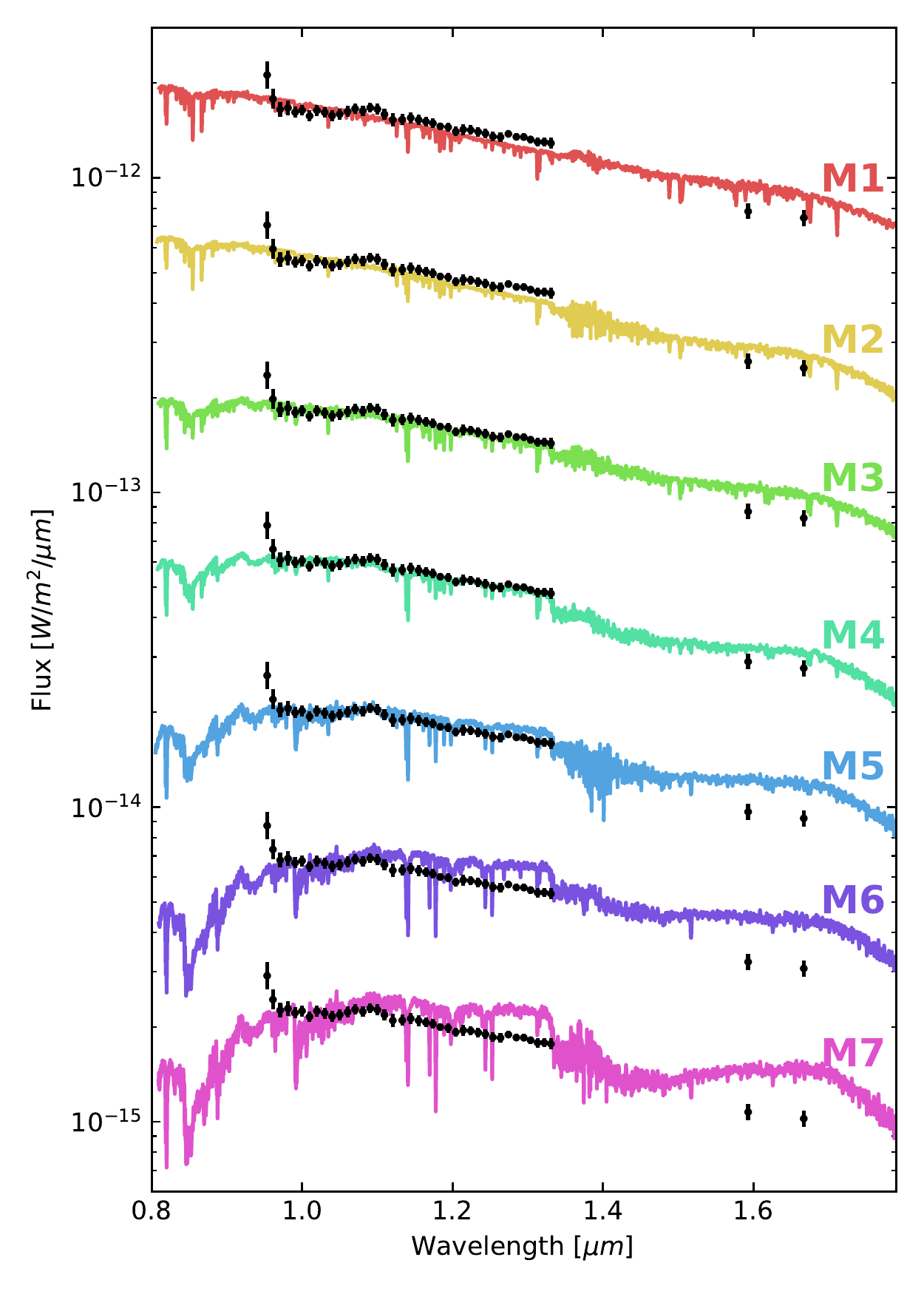}
\caption{Spectrum for HD~133778B. Both the IFS (YJ) and IRDIS (H) bands are shown. HD133778 shows a good match to spectral types $\sim$M3-M4, which is in agreement with the absolute magnitude of this object. Reference spectra are from \citet{cushing2005}, and the stars used as reference objects for M1 to M7 respectively are GL229A, GL411, GL388, GL213, GL51, GL406 and VB8.}
\label{fig:spectrum133778}
\end{figure}

The companion falls within the IFS field of view for a subset of frames, and is sufficiently bright that a low resolution spectrum (R$\sim$35) can be extracted even from just these few data frames. Since the total integration time is significantly shorter in the second observation, we only use the data from 2016-05-16 for the spectral and photometric analysis. The extracted spectrum, and the H2/H3 band photometry, are shown in Figure \ref{fig:spectrum133778} with M1-M7 reference spectra from \citet{cushing2005} also plotted. Visually, HD~133778~B matches the M3 and M4 spectra best, and the absolute magnitude is a good match for an M4 star (see Figure \ref{fig:cmd} above) so we class this object as M4$\pm$1.

HD~133778 is an old star, as described in Section \ref{sec:ages}. The object was previously a late A-type star, but has recently left the main sequence. It is unusual for such an old star to host a massive debris disk. The disk could either have been formed in a recent massive collision between small planets in the system, or between a planet and a planetesimal belt. Alternatively, the disk might have lasted the entire lifetime of the system. We discuss these scenarios in more detail in Section \ref{sec:discussion}

\subsection{HD~19257B: a newly identified stellar companion to an old early-F star}
\label{sec:19257}

We detected a bright companion to HD~19257, at a separation of 1.293$\pm$0.003'' and a position angle of 127.70$\pm$0.16$^\circ$. The object is clearly identifiable even in the flux calibration frames, and both the contrast and the astrometry of this candidate were measured from the flux calibration frames directly. This avoids the uncertainty in stellar flux and position introduced by the coronagraph, and the self subtraction introduced by the PCA algorithm, which is a significant effect for a companion as bright as this and would bias both the astrometry and the photometry. We find contrast ratios of $\Delta\textrm{H2}$=3.28$\pm$0.04 and $\Delta\textrm{H3}$=3.16$\pm$0.04 for the object.

The relative astrometry of the candidate is in agreement with common proper motion with the host star, as shown in Figure \ref{fig:cpm_all}, confirming that this is indeed a bound companion. The position of the candidate changes by 10mas relative to the host star, with the two measurements consistent at the 2$\sigma$ level. The slight change in measured position could correspond to a marginal detection of orbital motion (which could cause an on-sky motion of up to $\sim$13mas/yr, see Table \ref{tab:companions}), but we do not attempt to fit the orbit with only two datapoints.

The companion is sufficiently bright that it also appears in the \gaia~DR2 catalog, with a G magnitude 11.53$\pm$0.03 \citep{gaia_dr2}. However, the companion is only marginally detected by \gaia, and no parallax or proper motion are quoted in the catalog. Indeed, the companion is at the limit of the contrast sensitivity of \gaia: at the measured \gaia~separation of 1.253'', \citet{brandeker2019} find a 50\% detection fraction at a contrast of 4.63 and a 90\% detection fraction of 4.27; the contrast of HD~19257B is 4.53$\pm$0.03mag in the G band. Given that this is a marginal detection, the companion flux should perhaps be interpreted with caution.

Assuming that the \gaia~flux is correctly measured, we derive stellar properties for the companion using
\texttt{isochrones} \citep{mortonisochrones}. We input the \gaia~parallax of the host star HD19257~A and the G and H band apparent magnitudes of the companion. The median sample is a star with a mass of 0.54M$_\odot$ and radius 0.49R$_\odot$. We repeat the procedure using only the parallax and H band magnitude, and find a somewhat larger star with mass 0.69M$_\odot$ and radius 0.67R$_\odot$. When using only the H band, the median sample has a much bluer color than we measure (2.6 compared to 1.9), perhaps indicating an incorrectly measured flux in one of the bands. The \texttt{isochrones} package also predicts that this object is a very old star: whether or not the G-band magnitude is included, the median age of the samples is $\sim$7Gyr, which is significantly older than the age extracted for the host star of $\sim$800Myr, but likely unreliable due to the limited number of photometric measurements and the discrepancy between the predicted and measured G-H color of the object.

\begin{figure}[ht]
\centering
\includegraphics[height=0.46\textwidth, trim=0.5cm 5.2cm 3.6cm 5.2cm, clip]{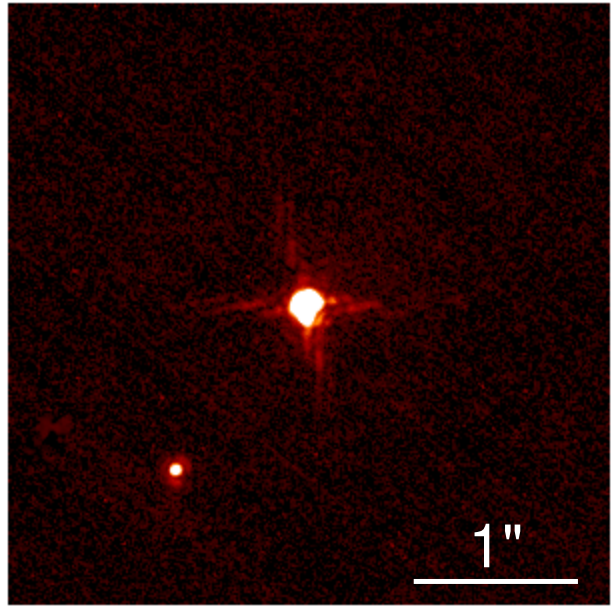}
\caption{The discovery image of HD~19257B. This is a flux calibration image taken with SPHERE/IRDIS in the H2 filter, and without the coronagraph present. Bad-pixels have been removed and North is oriented up. HD~19257A is visible in the center and HD~19257B to the lower left.}
\label{od:hd19257}
\end{figure}

\citet{fremat2007} recorded HD~19257 as a suspected binary due to discrepant radial velocity measurements, but the star is not identified as a spectroscopic binary in that work and the authors instead conclude that it is a variable star. That work quotes an effective temperature of 7367$\pm$195K and a log\textit{g} of 4.23$\pm$0.06. The target is listed as spectral type A5 by \citet{cannon1993}, and as an A9III by \citet{fremat2007}. Most recently, \citet{kuchner2016} determined a spectral type of F0V. This is consistent with \citet{mcdonald2012} and \citet{kennedywyatt2013} which do not quote spectral types but find effective temperatures of 7227~K and 7458~K respectively, and with \gaia~which finds a temperature of 7420$\pm$120K. Since the \gaia~catalog resolves the companion and finds a very similar temperature to \citet{fremat2007}, we conclude that these temperature fits are not affected by the presence of the companion.

We determine an age of 790$\pm$150Myr for this target. This is older than the majority of dust-hosting stars: the system is somewhat reminiscent of HD~133778, in that both are older stars, with stellar mass companions and large quantities of dust. Is it possible that these stellar companions are stirring planetesimals and maintaining the debris disk until much older than the typical age of highly dusty systems? Alternatively, perhaps planetary components of these systems are migrating due to the widely separated companion, and this migration has recently caused a ring of planetesimals to be disturbed, and elevated the dust volumes in these systems to detectable levels.

\subsection{HD~18378~B: a mid-M companion to a giant star}
\label{sec:18378}

\begin{figure*}[ht]
\centering
\includegraphics[width=0.6\textwidth, trim=1.8cm 1.8cm 1.8cm 1.4cm, clip]{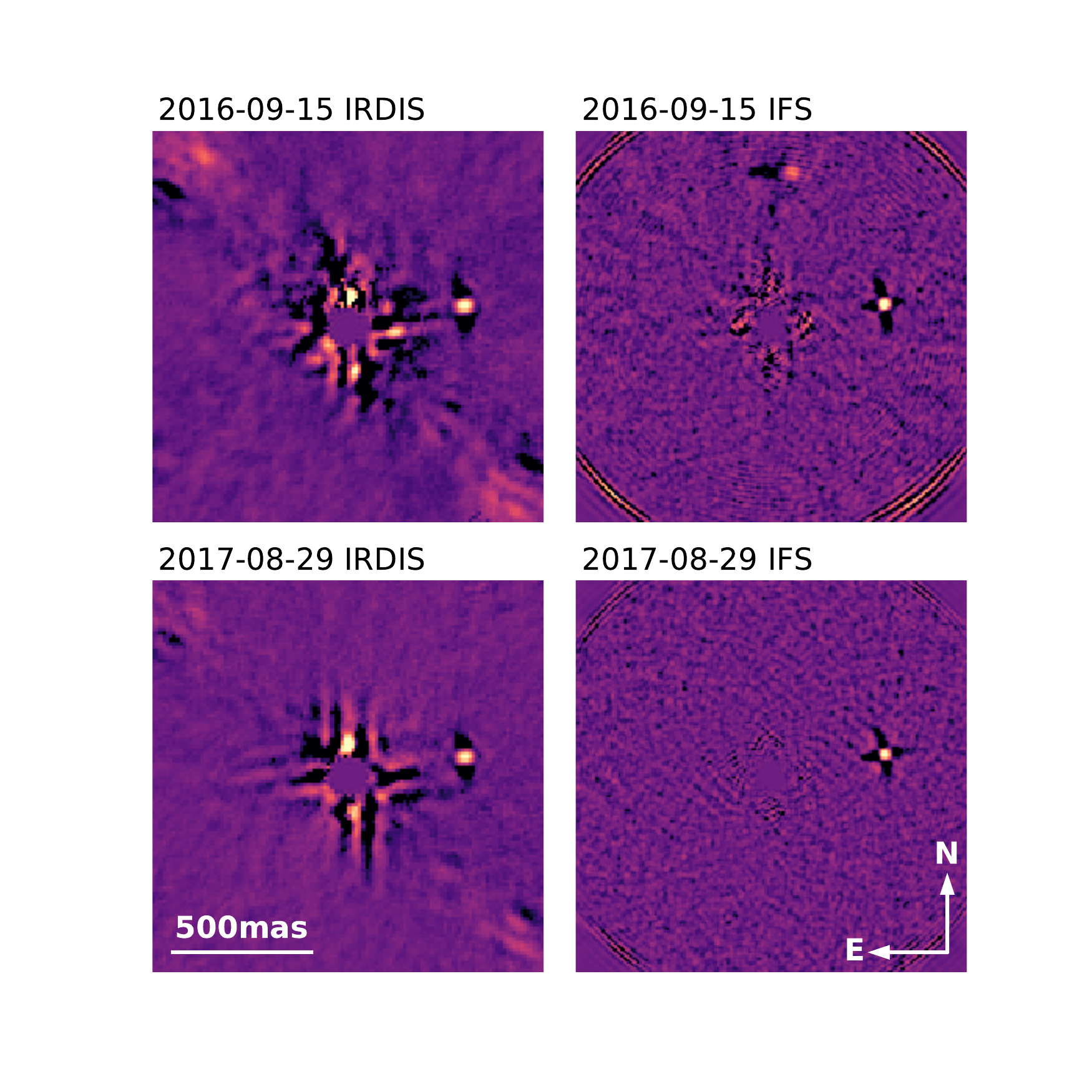}
\caption{Discovery images of HD~18378~B. Both IRDIS and IFS data are reduced using a PCA algorithm to remove stellar speckles, and only the central portion of the data is shown. The companion is clearly visible to the right of the star, which is behind an occulting mask in the center of the image. The point to the north of the star in the 2016-09-15 IFS data is due to flux persistence from a previous observation, and is not astrophysical in nature.}
\label{fig:hd18378}
\end{figure*}

We identify a companion to HD~18378 for the first time (Figure \ref{fig:hd18378}), at a separation of 418$\pm$3mas from the host, a position angle of 279.4$\pm$0.4$^\circ$ and a contrast of $\Delta$H2=10.76$\pm$0.05mag. The object shares common proper motion with the host (see Figure \ref{fig:cpm_all}), with the relative astrometry consistent between epochs. For a face-on, circular orbit, the orbital period of this companion would be $\sim$2600 years, corresponding to a movement of just $\sim$1mas.

\begin{figure}[ht]
\centering
\includegraphics[width=0.44\textwidth]{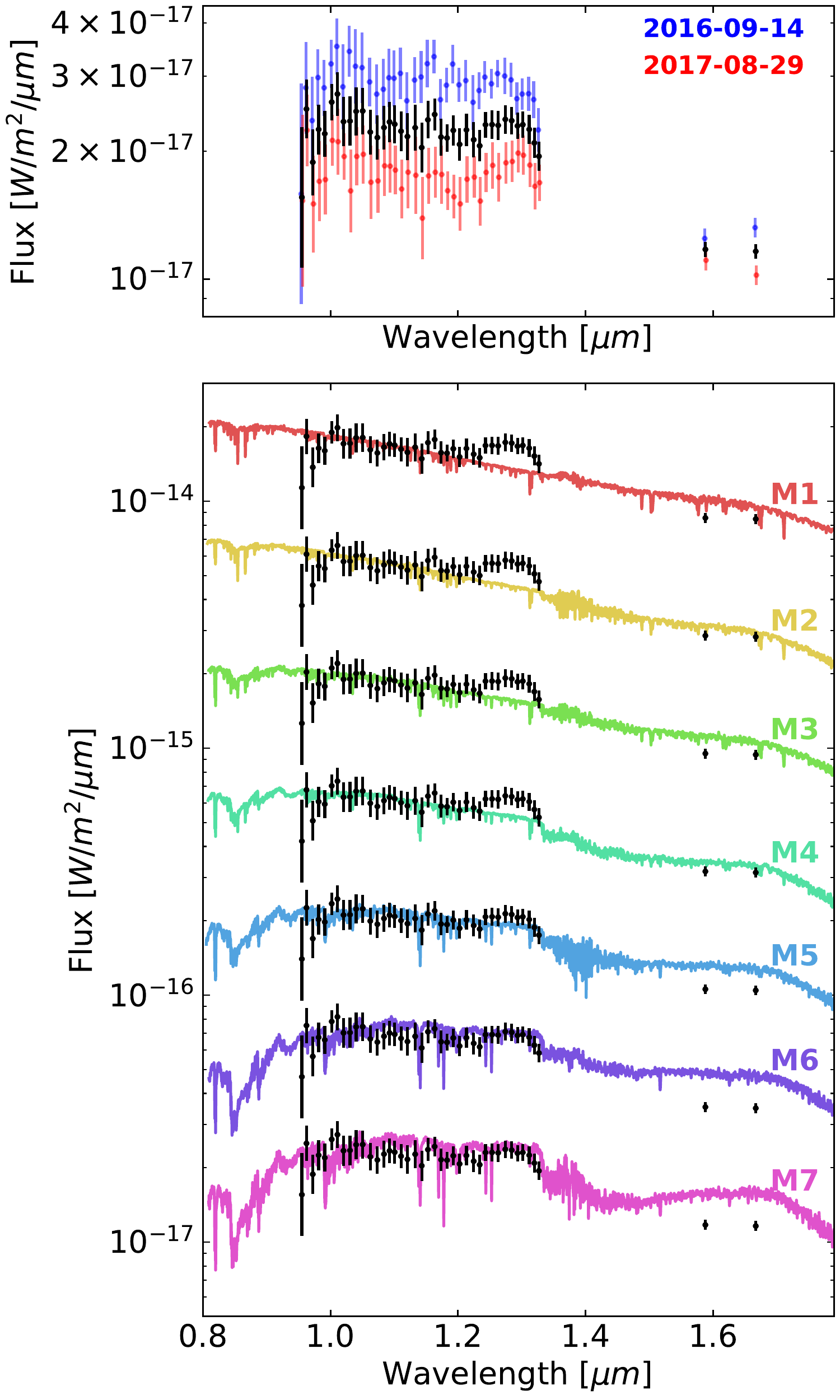}
\caption{{\it Upper:} Spectra for HD18378B as observed on 2016-09-14 (blue) and 2017-08-29 (red). Both the IFS (YJ) and IRDIS (H) bands are shown. The spectra show broadly similar shapes with an absolute offset, likely due to the flux calibration as detailed in the main text. In both cases, a peak is seen at $\sim$1.3$\mu$m, but this is much more pronounced in the second (red) spectrum. {\it Lower:} The median of the two spectra is plotted against seven reference spectra. Reference stars are from \citet{cushing2005} and are the same stars as those in Figure \ref{fig:spectrum133778}. Visually, the closest match is an early to mid M star, and using both the spectral and absolute magnitude information we type the object as M4$\pm$2.}
\label{fig:spectrum18378}
\end{figure}

The low-resolution spectrum of this object in the YJ band, and the two H band photometric points, are shown in Figure \ref{fig:spectrum18378}. Spectra were extracted for both epochs of observation, and are plotted against each-other. There is a clear absolute flux offset between the two epochs. This is likely a calibration offset due to weather changes within an observation: since the host star and the companion are observed concurrently and not simultaneously, slight changes in transmission of the atmosphere between the flux calibration and the science observation can cause such offsets. This calibration offset should not affect the shape of the spectra. For both epochs, the spectrum shows a relatively flat slope between 0.9-1.2$\mu$m, and a small peak between 1.2-1.4$\mu$m, which is significantly more pronounced in the second epoch.

In the second panel of Figure \ref{fig:spectrum18378}, the median of the two epochs is plotted against reference spectra for M1-M7 field dwarfs from \citet{cushing2005}. Although the spectrum is relatively noisy, the slope short of 1.2$\mu$m matches that of the reference spectra between M1 and M5, while the peak at 1.3$\mu$m is visually a closer match to spectral types M5-M6. None of the reference spectra accurately capture both the slope of the relatively flat 1.0-1.2$\mu$m region of the spectrum and the peak in emission between 1.25-1.3$\mu$m. The absolute flux of the object in the H band is similar to that of M4 field dwarfs (see Figure \ref{fig:cmd}). We therefore type this object as M4$\pm$2, with more data required for a more precise spectral classification.

Very limited literature exists for this target with the only references mentioning the host star being \citet{houk1975} and \citet{olsen1994}, who both list it as a K2III star. \gaia~DR2 quotes the star as having T$_\textrm{eff}=4590\pm80$K and R=11.9$\pm$0.4R$_\odot$ \citep{gaia_dr2} which appears to be in agreement with the classification of this star as a giant. 

There are two potential interpretations of this system. This star would have been an early A object while on the main sequence, and perhaps represents the later stages of evolution of some of the other stars in this survey. It could have maintained a debris disk throughout its lifetime, or perhaps a recent massive collision between planetesimals or even small planets in the system has elevated the rate of dust production to a detectable level. Alternatively, and perhaps more likely, it might be that we are simply observing infrared excess due to mass loss from this giant star. 

\subsection{Wide binaries identified in \gaia}
\subsubsection{HD98363B (Wray 15-788)}

HD98363 has previously been identified as being in a wide stellar binary with Wray 15-788, with the pair at a projected separation of $\sim$50'' ($\sim$6900au) \citep{bohn2019}. Wray 15-788 hosts a disk, with H$\alpha$ emission suggesting that the disk is protoplanetary \citep{pecautmamajek2016}. \citet{bohn2019} detected the disk in scattered light for the first time and discovered a large inner cavity, indicating that the disk is likely in the transition stage. This system is fascinating, since the HD98363 disk appears to be gas-poor and as such in the more evolutionarily advanced debris disk stage \citep{chen2012,moor2017}. Although resolving this disk would be highly interesting, it is somewhat unsurprising given the relatively low \Lir (see Discussion section) that we were not able to directly detect the HD98363 disk in this work. The disk has recently been resolved by \citet{hom2020}, using the GPI polarimetry mode \citep{perrin2010}, and that work discusses the architecture of the system in more detail.

\subsubsection{Other bound companions from \gaia}

We also search the \gaia~catalog for companion stars outside of the SPHERE field of view. For each star, we search for companions out to a projected separation of 10,000~au at the distance of the target. We select as potential companions those sources which have both parallaxes consistent with the host to 3$\sigma$, and proper motions in both RA and Dec consistent with the host to 8$\sigma$. This relatively wide constraint on proper motion is used to ensure that any differential motion between the two targets due to their mutual orbit does not exclude a bound pair from the selection. We recover three potential binaries: the HD98363+Wray 15-788 system discussed above, and anonymous field star companions to HD138564 and HD151012. These companions have \gaia~source IDs 6003218283764194304 and 6034203483517407616 respectively in the DR2 catalog.

To check that each pair is indeed a bound stellar binary, we estimate an approximate mass for each component, and calculate an approximate orbital velocity of the two stars relative to each other. This includes several coarse assumptions: we assume a circular orbit, and acknowledge that our estimates of secondary mass are very approximate, given that they are based only on the \gaia~photometric fits. These values can nonetheless be compared to the observed differential velocity of the two systems: a significantly higher differential velocity would indicate an unbound pair. A differential velocity below this threshold does not confirm that the pair is bound, since motion could be occurring along the line of sight and since we are using very approximate orbital values, but the combination of small separation, similar parallax and a low differential velocity is a good indicator that the system is a stellar binary. Both the HD98363 and HD151012 companions show differential velocities smaller than the expected orbital velocity; HD138564 and its companion shows a differential velocity marginally higher than the expected orbital velocity (2.24$\pm0.25$mas/yr compared to an expected value $\sim$1.95mas/yr). Given that the differential velocity is only marginally above this threshold, which is somewhat approximate, we conclude that all three pairs are likely stellar binaries.

The properties of all three of these companions are listed in Table \ref{tab:companions}. All three are in a similar parameter regime: the companions are at projected separations of approximately 6900au, 2300au and 3700au for HD98363, HD138564 and HD151012 respectively. Further, all three primary stars are members of the Scorpius-Centaurus association, and are therefore young. These are in a markedly different regime than the companions detected with SPHERE in this work, which are all within a few hundred au of their hosts and orbit older stars.


\section{Discussion}
\label{sec:discussion}

We identified three late-type stellar companions to highly dusty stars identified by \wise, one of which was also resolved by \gaia. We used \gaia~to confirm that three additional stars in the survey have equidistant and co-moving stellar companions at several thousand au, one of which had been previously identified by \citet{bohn2019}. The fraction of wide binaries ($\sim$30\%) appears to be in broad agreement with previous works: \citet{rodriguezzuckerman2012} find that $25\pm4\%$ of debris disks are in binary or triple star systems while \citet{thureau2014} found that the rates of debris disk occurrence around A-stars with and without stellar companions are similar ($26\pm7\%$ and $24\pm6\%$ respectively). These works both find a lack of binaries with intermediate separations between $1-100$~au, and the three SPHERE binaries presented here are consistent with that finding, with projected separations of 96.2$\pm$0.5au, 162$\pm$2au, and 218$\pm$3au. The actual separations in each case are likely higher, but cannot be accurately determined without additional astrometric monitoring of the candidates to determine their orbits. In each case, even the projected separation is significantly wider than the predicted dust radius, as is expected for non-destructive interference.

It is striking that all three objects with intermediate-separation stellar companions are relatively old: one is $\sim$790Myr, one has recently left the main sequence and one is a giant star. The post main sequence evolution of the star through the sub-giant and giant phases is relatively short lived, and as such it is less common to observe objects at this stage of evolution. For this survey to have detected multiple such stars with high dust volumes hints that it might be relatively common for these evolving systems to generate bright, but short-lived, debris disks. Even further, it seems that the combination of an older star with both a stellar companion and a bright debris disk might be common. 

One mechanism to explain these older systems with debris disk is the active migration of a planetary system, perhaps in the vicinity of a planetesimal belt. \citet{kratterperets2012} studied the effect of post main sequence stellar evolution on a planet in a binary star system, and found that stellar evolution can cause the planet orbit to evolve, possibly causing ejections, collisions, or star-hopping (meaning a planet orbiting an evolving primary moves into a chaotic orbit and then stabilizes in orbit around the secondary star). Although that work only studies single planets, we suggest that this complex and sometimes chaotic orbital evolution could also lead to collisions between a giant planet and additional planets in the system, or between a planet and a previously stable planetesimal belt. In either case, the collisions this process causes would lead to an elevated level of dust in the system for a short time, and a collision with a planetesimal belt in particular could trigger a collisional cascade that quickly generates large volumes of dust. Alternatively, chaotic orbital evolution to a very small separation from the star could lead to a tidal disruption event, which would also eject matter into the system and potentially trigger further collisions and thus dust production. Additional constraints on the composition and location of the disk could provide clues as to the true origin of this dust. Mid-IR spectroscopy may allow the radial location(s) of this dust belt to be constrained, and if the disks can be resolved in polarized light this would allow the dust location(s) and structure to be analyzed. Together, these measurements would provide clues as to whether this dust is indeed the product of a giant collision or tidal disruption event \citep[see e.g.][]{weinberger2011}, or is instead a debris disk that has survived the entire lifetime of the system \citep[similar to $\kappa$ CrB,][]{bonsor2013}. A ring of dust with a small radial extent, and a lack of cold dust at wider separations from the star, would be strong evidence that the dust in these systems was created in a recent dramatic collision, since in this case the dust could not have been replenished from a cold reservoir. Note however that the presence of cold dust does not preclude the possibility of the warm dust having been created in a recent collision.

No planets were detected in this survey, although this result is in line with the relatively low occurrence rate of planets accessible to high contrast imaging \citep[e.g.][]{meshkat2017,nielsen2019,vigan2020}. To assess the occurrence rates of planets in dusty systems will require a much larger sample size, likely only achievable through a meta-study of several campaigns that have prioritized highly dusty systems \citep[e.g.][]{janson2013,rameau2013,wahhaj2013,meshkat2015,matthews2018,launhardt2020}.

\begin{figure*}
\centering
\includegraphics[width=1.\textwidth]{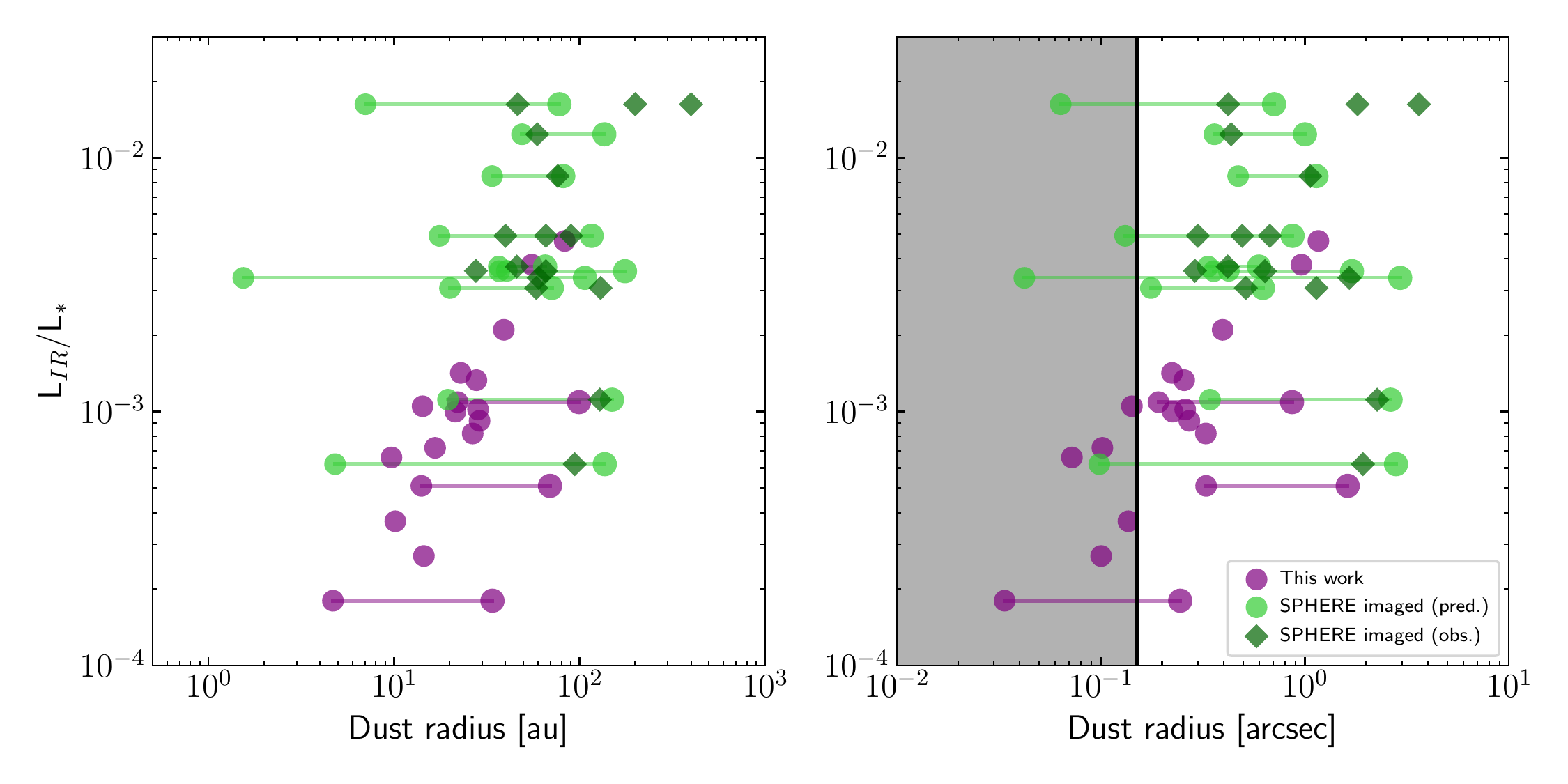}
\caption{Fractional luminosities and predicted radii of the debris disks in this survey (purple), and of those disks that have been imaged in scattered light with SPHERE in the H band (green). For each disk we use the blackbody temperature from \citet{cottensong2016} and convert the temperature to a predicted radius using the relations in \citet{pawellek2015}. Dust radii are shown in physical units on the left, and projected separation on the right, and for two-temperature disks we indicate both predicted radii, joined by a line. The inner working angle of \sphere~is additionally illustrated on the right with gray shading. Although most disks are outside the inner working angle of SPHERE, many are close and therefore likely obscured by the stellar speckles that are dominant at close separations. Many of the targets here are brighter than the faintest disks that have been imaged in scattered light with SPHERE, but are close to their stars and sufficiently faint that detection is very dependent on actual disk radius and orientation. We include only reference disks that have been imaged with \sphere~in the H-band using IRDIS, and note that some of these disks had previously identified with other platforms. For these disks, we also plot the radius (or radii) as measured from the resolved images using dark green diamonds. Reference data for resolved disks are from \citet{olofsson2016,lagrange2016,wahhaj2016,perrot2016,milli2017,choquet2017,matthews2017,bonnefoy2017,feldt2017,engler2019,gibbs2019}.}
\label{dustradii}
\end{figure*}

None of the disks in this survey were detected in scattered light. This is unsurprising given the small total sample size, and the typical temperatures and fractional luminosities of each disk. To demonstrate this point, we calculate the expected radii for each debris disk, based on the best-fit temperatures from \citet{cottensong2016}, for the 17 targets from our survey in that work. Temperatures are converted to radii following the process described in \citet{pawellek2015}, which takes into account the composition of dust grains, and we use the 50\% astrosilicate + 50\% ice composition. Temperature and radius values are given in Table \ref{tab:dustradii}.The dust radii and fractional infrared luminosities are plotted in Figure \ref{dustradii}, in both physical and angular separations. For comparison, we also show disks that have been imaged in scattered light with SPHERE/IRDIS in the H-band. Some of these had previously been imaged in scattered light with other platforms, while others were resolved for the first time with SPHERE. These disks are typically brighter than those in our survey, but several targets in our survey are of comparable brightness to some of the disks that have been imaged in scattered light. The GPIES disk survey \citep{esposito2020} found that only disks with excesses greater than \Lir=$10^{-4}$ were directly detected.

Although the disks in this survey are consistently above the \Lir=$10^{-4}$ threshold, several are within the inner working angle of the instrument. Some of those beyond this threshold are still within the parameter space where detections are hard, and very dependent on factors such as the alignment of the disk and disk composition. Of all the disks in the survey, HD176638 and HD182681 host the disks that are best suited to scattered light imaging: these systems have bright disks, widely separated from their stars and as such are in an ideal parameter space to be imaged in scattered light. Perhaps these disks are face on or close to face on, a configuration in which the integrated surface brightness is lower and in which the process of removing stellar speckles also strongly removes disk signal. If the disks are indeed face-on, polarimetric imaging might allow for them to be directly imaged.

\begin{table}[ht]
    \footnotesize
    \hspace*{-0.7cm}\begin{tabular}{lcccc}
        \hline
        Target & T$_1$ [K] & R$_1$ [au] & T$_2$ [K] & R$_2$ [au] \\
        \hline \hline
		HD19257    & 235 & 10 & --- & --- \\
		HD24966    & 170 & 22 &  80 & 100 \\
		HD94893    & 135 & 29 & --- & --- \\
		HD98363    & 350 &  5 & 130 &  34 \\
		HD113902   & 230 & 14 & --- & --- \\
		HD119152   & 140 & 27 & --- & --- \\
		HD123247   & 130 & 39 & --- & --- \\
		HD133778   & 190 & 17 & --- & --- \\
		HD138564   & 175 & 23 & --- & --- \\
		HD138923   & 275 & 10 & --- & --- \\
		HD151029   & 140 & 28 & --- & --- \\
		HD151012   & 150 & 28 & --- & --- \\
		HD157728   & 200 & 14 &  90 &  69 \\
		HD158815   & 170 & 15 & --- & --- \\
		HD176638   & 115 & 55 & --- & --- \\
		HD182681   &  90 & 83 & --- & --- \\
		HD207204   & 170 & 21 & --- & --- \\
        \hline
    \end{tabular}
    \caption{Predicted dust temperatures and radii. We quote temperatures from \citep{cottensong2016}, and list here only the 17 targets from our survey that are also in that work. Temperatures are converted to radii following the relation from \citet{pawellek2015}, and we use their ``50\% astrosilicate+50\% ice'' grain composition. Where the best fit from \citep{cottensong2016} suggests the presence of two dust belts, we list both temperatures and the corresponding radii for each.}
    \label{tab:dustradii}
\end{table}


\section{Conclusions}
\label{sec:conclusion}

We have presented a survey of 20 systems with mid-infrared excesses as detected by \wise. All of these stars have been carefully vetted to confirm that the excess is real and associated with the star, rather than being contamination from a background source. 19 of these are nearby, early-type, debris disk hosting stars, while one of the objects is a giant star, and so the excess could instead be due to mass loss. 

We detected three stellar companions in our \sphere~observations, and found evidence for three additional stellar companion in \gaia, which were outside the field of view of the current observations. We collected spectroscopy of two of these stellar companions in the YJ band, and by comparing these to the spectra of field stars found that both are $\sim$M4 type. We present preliminary estimates of the likely orbital parameters of these systems, based on the projected separations and companion spectral types, but do not attempt orbit fitting given that we have only two astrometric measurements of each over a relatively short time baseline. Further astrometric monitoring would allow the true separation, eccentricity and orbital period of these companions to be extracted. We also detected three more widely separated companions in the \gaia~catalog, a result that seems broadly in line with the expected stellar companion rate for debris disk stars.

The systems with intermediate-separation stellar companions are all old. We suggest that these are similar to other late-type stars with large volumes of dust, such as BD+20 307, where \citet{weinberger2011} inferred that the dust was likely a result of a terrestrial planet being destabilized from its orbit and causing a massive collision. Mid-IR spectroscopy or polarimetric imaging of the post-main sequence dusty systems presented here could confirm the disk location and spatial extent, and thus demonstrate whether the dust is indeed the result of a recent collision, or if it is instead an unusually long-lived dust disk.

Due to the small sample size of this project, we do not perform a statistical analysis of planet occurrence rates in debris systems. A future work should combine results from several surveys studying debris disk excess stars to confirm with high statistical confidence whether exoplanet companions are indeed more common in the dustiest systems, as has been hinted at in \citet{meshkat2017}.


\acknowledgements
{
We thank the anonymous referee for their thorough reading of this work, which improved the quality of the manuscript.
AV acknowledges funding from the European Research Council (ERC) under the European Union's Horizon 2020 research and innovation programme (grant agreement No. 757561). Part of this research was carried out at the Jet Propulsion Laboratory, California Institute of Technology, under a contract with the National Aeronautics and Space Administration.
Based on observations collected at the European Organisation for Astronomical Research in the Southern Hemisphere under ESO programmes 095.C-0426(A), 097.C-1042(A) and 099.C-0734(A).
This research has made use of the VizieR catalogue access tool, CDS, Strasbourg, France.  The original description of the VizieR service was published in A\&AS 143, 23. This research has made use of the SIMBAD database, operated at CDS, Strasbourg, France.
}

\facilities{VLT:Melipal,WISE,Gaia}

\software{astropy \citep{astropy},
isochrones \citep{mortonisochrones},
matplotlib \citep{matplotlib}, 
numpy \citep{numpy},
pandas \citep{pandas}
}
	

\bibliographystyle{aasjournal}
\bibliography{matthews_arxiv_bib}


\appendix

\section{Target Notes}
\label{appendix:targetnotes}

This appendix describes the candidate companions identified for each individual target within this survey, and our determination of each candidate as a companion or a background star.

\textbf{HD~18378:} We detect a co-moving companion (see Section \ref{sec:18378}) and no other candidates.

\textbf{HD~19257:} A single, bright co-moving companion is identified, as discussed in Section \ref{sec:19257}.

\textbf{HD~24966:} No candidate companions are detected for this target.

\textbf{HD~94893:} Five candidates are detected. The median movement of the candidates between our two data epochs is a little less than expected, perhaps indicating a small mis-calibration of the stellar position in one of the two observations. Nonetheless, each object moves significantly between the two epochs, and we conclude that they are all background objects.

\textbf{HD~98363:} This target was observed in poor weather in 2015, and six candidates detected. A follow-up observation in significantly better weather conditions was carried out in 2018. The six candidates from the 2015 observation were all recovered and confirmed to be backgrounds based on their proper motion, and an additional 13 candidates were detected. Eleven of these are faint (M$_{H2}<$15) and have H2-H3$\sim$0, and we conclude that these are also background stars. Of the remaining 2, one is too bright for H2-H3 color to differentiate companions and background stars, and one is offscreen in the H2 filter. Both are likely background objects given their separation from the host, but more data would be required to confirm this conclusion.

\textbf{HD~113902:} Eight candidates are detected, and all show good agreement with the background hypothesis between our two epochs of data.

\textbf{HD~119152:} No candidate companions are detected for this target.

\textbf{HD~112802:} One wide separation candidate is marginally detected on the very edge of the detector, at a separation of $6.3\arcsec$ from the host star. Both astrometry and photometry are biased this far from the target star, and so we do not include this candidate in our work.

\textbf{HD~123247:} Six candidates are identified in two epochs of data, and common proper motion testing confirms that all of these candidates are background objects.

\textbf{HD~133778:} Two candidates are detected, at $0.99\arcsec$ and $5.2\arcsec$. The close candidate is confirmed to be co-moving, and is discussed in detail in Section \ref{sec:133778}, while the object at 5.2$\arcsec$ does not share common proper motion with the target and is a background object.

\textbf{HD~138564:} Four candidates are detected, all of which show consistent motion with background objects.

\textbf{HD~138923:} A single candidate is detected in one epoch of data. The candidate is sufficiently bright that CMD analysis does not allow companions and background stars to be differentiated. Further work is required to confirm if this object is a companion, although at a projected separation of 3.98$\arcsec$ (534~au), it seems unlikely that the object is bound.

\textbf{HD~151012:} This target is observed twice: twelve candidates are found in the first epoch and nine are recovered in a second epoch of data and confirmed to be background objects. The remaining three objects are confirmed to be background stars based on their CMD position.

\textbf{HD~151029:} No candidate companions are detected for this target.

\textbf{HD~157728:} No candidate companions are detected for this target.

\textbf{HD~158815:} This object is observed twice and 98 candidates are detected: 86 in both epochs, 10 only in the first epoch and 2 only in the second epoch. All of the candidates that are detected twice are clear background objects (see Figure \ref{fig:cpm_all}). A further 3 objects are likely backgrounds given their position on a CMD (see Section \ref{sec:cmd}). Finally, seven candidates are sufficiently bright that the H2-H3 color does not distinguish companions and background stars, and two candidates are only in the field of view in the H3 images, and so we cannot calculate an H2 magnitude or an H2-H3 color. Many of these candidates for which color is not a determinant are very close to the conservative threshold that we use to confirm background position from CMD position alone, and are still likely to be background objects. These ambiguous objects are also all separated from the host by at least 2'', with all but two separated by more than 5'',  and the field is particularly crowded, so we suggest that these are all highly likely to be background sources.

\textbf{HD~176638:} Three candidate companions are detected, and all are confirmed to be background objects based on astrometry at two epochs of data.

\textbf{HD~182681:} One candidate is detected, and the target is only observed once so we cannot test for common proper motion. The candidate is too bright for the H2-H3 color to indicate whether this object is likely to be a genuine companion.

\textbf{HD~192544:} No candidate companions are detected for this target.

\textbf{HD~207204:} No candidate companions are detected for this target.

\end{document}